\documentclass[aps,pre,reprint,amsmath,amssymb,superscriptaddress,showpacs,floatfix]{revtex4-2}

\usepackage{graphicx}
\usepackage{dcolumn}
\usepackage{bm}
\usepackage[colorlinks, allcolors=blue]{hyperref}
\usepackage[usenames, dvipsnames]{color}
\usepackage{float}

\usepackage{lipsum}
\usepackage{titlesec}
\titlespacing\section{0pt}{12pt plus 4pt minus 4pt}{1pt plus 20pt minus 2pt}
\usepackage{xcolor}
\usepackage{tabularx}
\usepackage{amsmath}
\usepackage{comment}
\usepackage{afterpage}
\usepackage{placeins}
\usepackage{booktabs}
\usepackage{multirow}
\usepackage{array}
\usepackage{setspace}
\graphicspath{{Figs/}}
\usepackage{siunitx}
\usepackage{hhline}
\usepackage{xfrac}
\usepackage{float,graphicx}
\usepackage{mathtools}
\usepackage{listings}
\usepackage{amssymb}
\usepackage[normalem]{ulem}
\usepackage{titlesec}
\usepackage{amsfonts}
\usepackage[version=4]{mhchem}

\usepackage{epstopdf}
\catcode`@11
\def\seceqaa{\@addtoreset{equation}{section}
principles\def\theequation{A\arabic{equation}}}
\def\seceqbb{\@addtoreset{equation}{section}
\def\theequation{B\arabic{equation}}}
\def\seceqcc{\@addtoreset{equation}{section}
\def\theequation{C\arabic{equation}}}
\def\seceqdd{\@addtoreset{equation}{section}
\def\theequation{D\arabic{equation}}}
\def\seceqee{\@addtoreset{equation}{section}
\def\theequation{E\arabic{equation}}}
\def\seceqff{\@addtoreset{equation}{section}
\def\theequation{F\arabic{equation}}}
\def\seceqgg{\@addtoreset{equation}{section}
\def\theequation{G\arabic{equation}}}
\def\seceqhh{\@addtoreset{equation}{section}
\def\theequation{H\arabic{equation}}}
\catcode`@11

\DeclareUnicodeCharacter{2009}{\,}

\newcommand{\SPINX}{\affiliation{Spin-X Institute, School of Physics and Optoelectronics, State Key Laboratory of Luminescent Materials and Devices, Guangdong-Hong Kong-Macao Joint Laboratory of Optoelectronic and Magnetic Functional Materials, South China University of Technology, Guangzhou 511442, China}}
  
\newcommand{\IACS}{\affiliation{School of Physical Sciences, Indian Association for the Cultivation of Science, Jadavpur, Kolkata 700032, India}}

\begin{document}

\title{Revisiting MnSe : a Magnetic Semiconductor with Spin-Phonon coupling} 
\author{Suman Kalyan Pradhan}
\email{suman1kalyan@scut.edu.cn}
\SPINX

\author{Arnab Bera}
\IACS

\author{Soham Das}
\IACS

\author{Yongli Yu}
\SPINX

\author{Jicheng Wang}
\SPINX


\author{Rui Wu}
\email{ruiwu001@scut.edu.cn}
\SPINX

\begin{abstract}
Spin-phonon interactions in 2D magnetic materials are crucial in advancing next-generation spintronic devices. Therefore, identifying new materials with significant spin-phonon interactions is of great importance. In this context, MnSe, previously recognized as an exemplary non-layered $p$-type semiconductor emerges in this study as an intriguing material with notable spin-phonon characteristics. The complex magnetism in pristine MnSe, primarily dominated by antiferromagnetism with a weak ferromagnetic component, gives rise to both spontaneous and conventional exchange bias effects at low temperatures. In an effort to understand this intriguing magnetism, we conducted a detailed Raman spectroscopy study, 
which reveals unconventional deviations from the usual phonon anharmonicity around Néel temperature (170 K), in the self-energies of the P$_1$ ($\sim$ 233 cm$^{-1}$), P$_2$ ($\sim$ 256 cm$^{-1}$), and P$_3$ ($\sim$ 470 cm$^{-1}$) modes. Notably, the P$_1$ mode is most sensitive to spin-phonon coupling, while the P$_2$ mode is particularly responsive to the structural phase transition at 250 K. Therefore, these findings provide comprehensive insights into the phase transitions of pristine MnSe, particularly highlighting the previously unobserved interplay between its magnetic behavior and phonon dynamics. 
\end{abstract}
\maketitle
 \begin{figure}
\centering
\includegraphics[width=1\columnwidth]{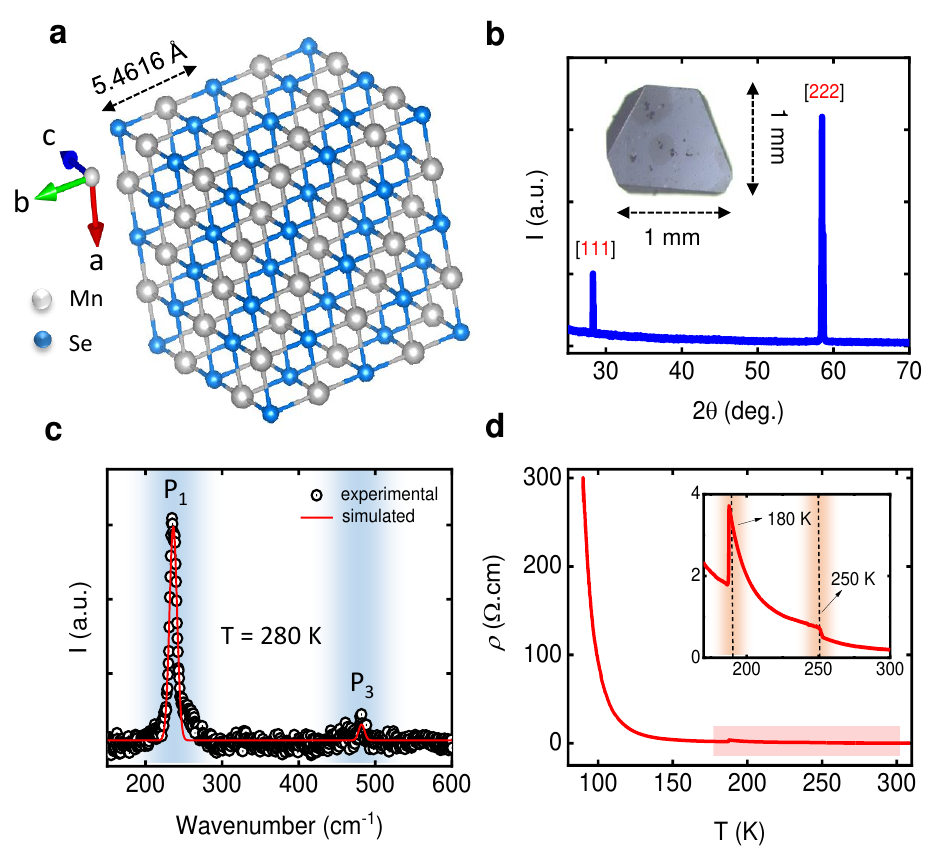}
\caption{\textbf{Initial characterization :}  (a) Schematic of the crystal structure obtained from room-temperature SC-XRD analysis, shown in a top view. The arrow indicates the unit cell thickness, with Mn and Se atoms represented by gray and cyan spheres, respectively. 
(b) XRD pattern of MnSe single crystals. The inset displays a photographic image of a grown crystal. (c) Representative Raman spectrum at 280 K in the range of 150–600 cm$^{-1}$. The open circles and solid lines correspond to the experimental and simulated results, respectively. (d) Temperature-dependent resistivity plot indicating a semiconducting nature throughout the scan. The inset highlights two anomalies around 180 K and 250 K, associated with magnetic ordering and a structural phase transition, respectively (discussed in detail below).}
\vspace{-0.45cm}
\label{charect}
\end{figure}
\section{Introduction}


Semiconductors, with tunable electrical transport, and magnetic materials, with tunable spin configurations and magnetizations, form the foundation of modern-day information technologies. A long-standing challenge has been to develop materials that seamlessly integrate and link these two distinct properties. Magnetic Semiconductors are a key class of materials that provide a unique opportunity to explore the interplay of charge, spin, and lattice in distinct energy scales, leading to a wide range of electronic and magnetic phenomena \cite{Haas,PhysRevLett.88.207208,Wilson2021}. When spin interactions are strong, the formation of Magnetic Semiconductors can significantly enhance their functionality, making them highly suitable for use in spintronic devices. \cite{Wolf2001,Awschalom}.

Manganese selenide (MnSe), a non-layered $p$-type semiconductor, is one such example, which exists in three crystal structures: Rock Salt ($\alpha$), Zinc Blende ($\beta$), and Wurtzite ($\gamma$) \cite{Schlesinger1998}. The intricate nature of these structural phases is closely tied to their resulting physical properties \cite{PhysRevB.46.12076, RAO1976207}. Phase $\alpha$ (cubic) is thermodynamically stable at room temperature under normal atmospheric pressure \cite{EfremDSa2004}, and undergoes a partial transition to phase $\beta$ (hexagonal) below 270 K \cite{EfremDSa2004, Huang2019}. In contrast, the Wurtzite phase in MnSe is exceedingly rare \cite{https://doi.org/10.1002/anie.201001213}. Numerous studies report that MnSe is primarily an antiferromagnetic (AFM) material, with the ordering temperature varying between 120 K and 197 K \cite{Squire1939, Lindsay1951, Ito1978, EfremDSa2004, Popoviifmmodeacutecelsecfi2006, Huang2019}, depending on the synthesis method. Moreover, MnSe demonstrates a diverse array of intriguing properties across magnetism \cite{O’Hara2018, Hung2021}, structure \cite{Wang2016a}, electronics \cite{Prasad1991, Lei2006}, optoelectronics \cite{Sahoo2018}, and magnon scattering \cite{PhysRevB.66.012302}. A number of studies on various applications \cite{Raman2019, Miao2021, ZuchengZhang2021, Park2022} highlight its potential, making MnSe a compelling subject for investigating the intricate functional properties linked to its structural characteristics.

To date, MnSe has primarily been synthesized using the solid-state reaction method \cite{Huang2019} and various chemical vapor deposition (CVD) techniques \cite{Zhang2021,Zhou2022,Zhu2023}. 
However, it is important to note that MnSe synthesized via CVD often suffers from the common issue of poor uniformity. As a result, despite extensive studies on MnSe synthesized through different methods, obtaining high-quality single crystals has remained a challenge.

This research aims to address the long-standing challenge of synthesizing single-crystalline MnSe and explore its magnetic, transport, and phonon properties. 
High-quality, hexagonal-shaped, mm-sized single crystals [see the inset of Fig.~\ref{charect}(b)] were successfully synthesized using the chemical vapor transport (CVT) method. The Rock-Salt (cubic) crystal structure was confirmed through single-crystal X-ray diffraction (SC-XRD) experiments. Magnetic measurements reveal that MnSe undergoes paramagnetic to antiferromagnetic (AFM) transitions at around 170 K, along [111]. Notably, the antiferromagnetic order coexists with a weak ferromagnetic (FM) component, as evidenced by the prominent exchange bias effect. Another transition is observed around 250 K, corresponding to a partial structural phase transition ($\alpha \rightarrow \beta$) in MnSe. A significantly low inverse magnetocaloric effect (IMCE) is about 44 $\times  10^{-4}$ J/Kg-K, observed near the $T_\text{N}$. Electrical resistivity measurements demonstrate overall semiconducting behavior across the measured temperature range, with anomalies linked to magnetic and structural transitions. To uncover the underlying mechanisms, we present an experimental Raman scattering study focusing on the phonon properties within the temperature range of 100 to 300 K. 
A change in Raman intensity observed near 170 K, can be considered as a fingerprint of the magnetic transition. The temperature-dependent Raman study reveals unconventional deviations from the usual phonon anharmonicity in all observed Raman modes (P$_1$, P$_2$, and P$_3$). Notably, the P$_2$ mode is sensitive to the structural transition around 250 K, while the P$_1$ mode exhibits the strongest response to spin-phonon coupling. 
These observations position MnSe as a promising 2D material with intriguing functional properties.

The article is organized as follows. Section II presents all experimental results related to the pristine composition, along with a discussion on the relationship between spin and phonon attributes. Finally, Section III concludes this article.

\section{Results and Discussions}

\subsection{Characterization}

Single crystal samples of MnSe were grown by
using the chemical vapor transport method. The details of the synthesis method are described in the S1 section of Supporting Information. The optical microscopy image of a grown crystal is displayed in the inset of Fig.~\ref{charect}(b). For structure determination, we performed SC-XRD at room temperature of a suitable crystal. MnSe possesses a 
NaCl-type (Rock-Salt) structure, and crystallizes in ${Fm}$$\overline{3}$$m$ (225) and point group $O_\text{h}$ with the lattice parameter of $a$ = 5.4616 $\mathring{A}$. Table SI provides refined lattice parameters and atomic positions in detail. Fig.~\ref{charect}(a) displays
the top views of the refined crystal structure of MnSe, where Mn and Se
atoms are arranged alternatively and connected by strong equidistance chemical bonds. This further rules out the possibility of a layered structure.
Fig.~\ref{charect}(b) shows the typical XRD pattern of MnSe single crystal with intensity peaks in the [111] and [222] Bragg planes, indicating
that the crystal growth plane is along the 111 direction. As shown in Fig. S1, quantitative analysis of the Energy dispersive X-ray spectroscopy (EDXS) spectrum reveals an atomic ratio of Mn to Se in the as-synthesized sample of 1:1, consistent with the stoichiometric ratio of MnSe. 

A representative Raman spectrum of MnSe at $T = 280$ K is shown in Fig.~\ref{charect}(e). Two prominent Raman peaks are observed around 233 cm$^{-1}$ and 470 cm$^{-1}$. These modes are in excellent agreement with previous studies \cite{milutinovic2004,PhysRevB.73.155203} and are attributed to the longitudinal optical (LO) mode and the second harmonic of a one-magnon optical excitation (2M$_\text{OPT}$), respectively. In this symmetry, the second-rank tensor decomposes into three irreducible components: $A_\text{1g}$, $T_\text{2g}$, and $E_\text{g}$.

At zero magnetic field, the resistivity increases with decreasing temperature in the range of 90–310 K, exhibiting semiconducting behavior. Notably, two distinct anomalies appear at $T = 180$ K and 250 K, which are attributed to magnetic and structural phase transitions, respectively \cite{Ito1978}, as discussed later. 


\subsection{Magnetic properties of MnSe}
\begin{figure}
\centering
\includegraphics[width=1.00\columnwidth]{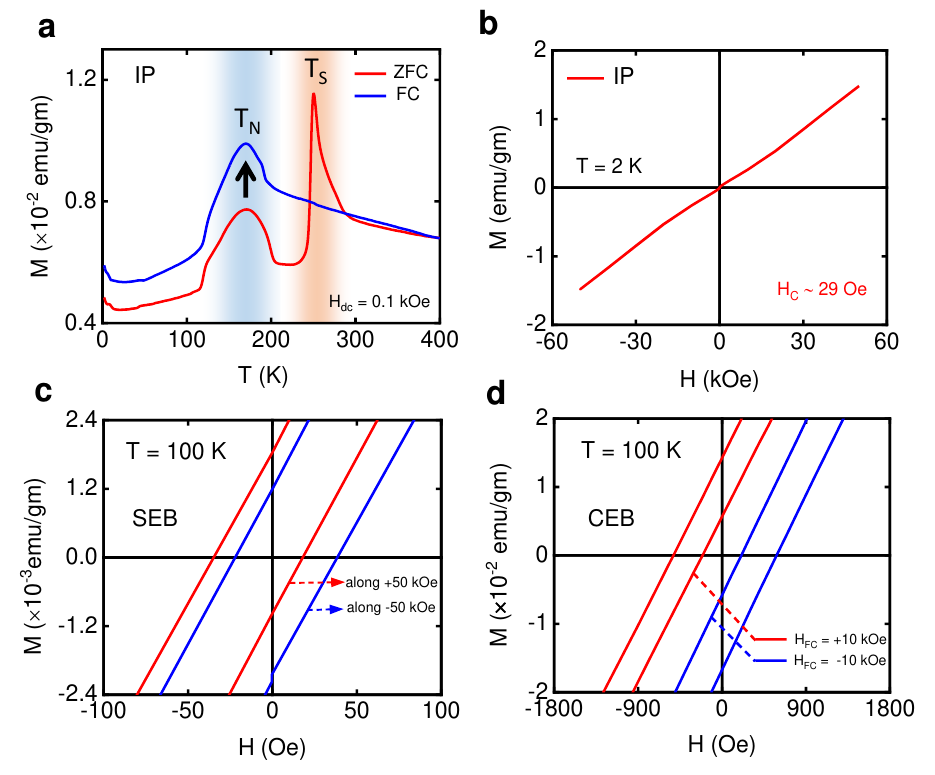}
\caption{\textbf{DC magnetic characterization :} Temperature-dependent magnetization [$M(T)$] measured under ZFC and FC modes with a magnetic field $H_\text{dc} = 0.1$ kOe applied along the IP direction. ZFC and FC data are represented by red and blue solid lines, respectively. 
(b)  Isothermal magnetization for the IP configuration, measured at $T = 2$ K, up to $H = 50$ kOe. The corresponding coercivity is indicated in the lower-right inset. (c) Spontaneous Exchange Bias (SEB): ZFC $M(H)$ loops at 100 K for the studied sample, measured following the sequence 0 $\rightarrow$ 50 $\rightarrow$ - 50 $\rightarrow$ 50 kOe (blue) and 0 $\rightarrow$ - 50 $\rightarrow$ 50 $\rightarrow$ - 50 kOe (red). (d) Conventional Exchange Bias (CEB): $M(H)$ loops at 100 K measured after field cooling in the presence of $\pm$10 kOe. These two measurement protocols confirm the presence of both spontaneous and conventional exchange bias.}
\vspace{-0.45cm}
\label{M-T-H}
\end{figure}

The temperature dependence of the DC magnetization from 400 to 2 K was measured using both zero-field-cooled (ZFC) and field-cooled (FC) protocols under various applied magnetic fields. Fig.~\ref{M-T-H}(a) shows the $M(T)$ data collected at an external magnetic field of 0.1 kOe along the in-plane (IP $\rightarrow$ $H \parallel$ 111) direction (see the Supporting Information section for $M(T)$ measured at $H_\text{dc}$ = 1 and 10 kOe). Two distinct anomalies are observed in the scanned temperature range: one between 120 K and 200 K, and another between 250 K and 300 K. The high-temperature transition ($T_\text{S}$ = 250 K) is associated with the partial transformation from cubic to hexagonal phase ($\alpha \rightarrow \beta$) \cite{EfremDSa2004,Huang2019}, while the low-temperature anomaly, primarily a paramagnetic $\rightarrow$ antiferromagnetic transition ($T_\text{N}$ = 170 K), is attributed to the coupling effects of the magnetic locking in $\beta$-MnSe and thermal fluctuations affecting the short-range ferromagnetic sheets in $\alpha$-MnSe \cite{EfremDSa2004,Huang2019}. Notably, with increasing $H_\text{dc}$, the positions of both transitions remain nearly unchanged [see Fig. S2(a,b)].

As shown in Fig. S2(c), at high temperatures, the inverse susceptibility 
($\chi$$^{-1}$ = $\frac{H}{M}$) measured at 
$H_\text{dc}$ = 10 kOe for 
IP 
configuration is linear and follows the ${Curie}$–${Weiss}$ (CW) formula, 
$\chi$ = ${C/(T-\theta)}$ [$C$ = Curie constant = $\frac{\mu^2_\text{eff}}{8}$, $\mu_\text{eff}$ = effective paramagnetic moment, and $\theta$ = Curie-Weiss temperature]. A linear fit in this high-temperature region yields effective paramagnetic moments of $\mu$$_\text{eff}$ = 5.70 $\mu$$_\text{B}$, 
which is very close to the theoretical value for the high-spin state of Mn$^{+2}$, 5.916 $\mu$$_\text{B}$ \cite{Kittel2015}) and the Curie-Weiss temperature $\theta$ = - 371 K 
\cite{Kittel2015}). 
The negative value of $\theta$ further confirms the dominant antiferromagnetic exchange interaction between adjacent Mn$^{2+}$ ions mediated by Se$^{2-}$ bridges within the paramagnetic matrix of MnSe.

Some earlier studies have suggested the coexistence of a hidden ferromagnetic (FM) phase alongside long-range antiferromagnetic order in pristine MnSe \cite{EfremDSa2004,Huang2019}. To investigate whether a FM component develops in pristine AFM MnSe, we measured magnetic field-dependent magnetization $M(H)$ at several selected temperatures between 2 K and 300 K in zero-field cooling mode. 
As shown in Fig. \ref{M-T-H}(b), ZFC $M(H)$ curve at $T = 2$ K exhibits a small hysteresis loop and non-saturating behavior even at a magnetic field of 50 kOe, indicating the dominance of AFM interaction. However, the loop is non-linear at low fields, suggesting a weak FM background in the pristine AFM state. 


To further elucidate the evolution of the weak FM component, we plotted the coercivity ($H_\text{C}$) as a function of temperature [derived from the ZFC $M(H)$ loops at different temperatures] in Fig. S2(d). A sudden increase in coercivity is observed around 150 K 
along IP. 
Furthermore, Fig. \ref{M-T-H}(b) illustrates that the $M$($H$) curve in ZFC mode displays asymmetry and a shift along the negative field axis. This type of asymmetry and shifting in the hysteresis loops is characteristic of the exchange bias (EB) effect. To estimate the values of the EB field ($H_\text{EB}$) and the coercive field ($H_\text{C}$), we used the relations 
$H_\text{EB}$ = $\frac{H_+ + H_-}{2}$, and $H_\text{C}$ = $\frac{H_+-H_-}{2}$, where $H_\text{+}$ and $H_\text{-}$ are the cutoff fields ($M$ = 0 value) at the positive and negative field axes of the hysteresis loop, respectively.


\subsubsection{Exchange Bias effect}
To establish the spontaneous exchange bias (SEB) behavior of the studied sample, we conducted field-dependent magnetization $M(H)$ measurements at 100 K by scanning the initial magnetization curve in the - 50 kOe direction, as illustrated in Fig. \ref{M-T-H}(c). 
An enlarged view of the two $M(H)$ curves as shown in Fig. \ref{M-T-H}(c), corresponds to initial magnetization in both + 50 kOe and - 50 kOe directions, revealing a slight difference between them confirming the SEB phenomena. The SEB effect in MnSe arises from the coexistence of significant ferromagnetic (FM) and antiferromagnetic (AFM) components. Thus, we conclude that the highlighted SEB effect is a fundamental and intrinsic property of MnSe, with the initial magnetization curve likely serving as a bias field in this context \cite{Pradhan2018}. Consequently, in addition to the pristine AFM behavior, weak FM characteristics are also present in MnSe. The temperature dependence of the calculated SEB field ($H_\text{SEB}$) is shown in Fig. S2(d), where its magnitude gradually increases with temperature and reaches a maximum below $T_\text{N}$, similar to $H_\text{C}$.

In certain instances, materials display the conventional exchange bias (CEB) effect alongside the SEB phenomenon \cite{Pradhan2018}. To investigate the CEB of the studied sample, we conducted a hysteresis loop measurement at $T$ = 100 K, under field-cooled conditions. For this measurement, the sample was first cooled to 100 K with a $\pm$ 10 kOe bias field from 300 K, which is well above the antiferromagnetic transition temperature ($T_\text{N}$). The enlarged $M$($H$) curve within $\pm$
1800 Oe at 100 K is shown in Fig.~\ref{M-T-H}(d). A shift in the $M$($H$) loops in the opposite direction of the applied bias field, as depicted in Fig.~\ref{M-T-H}(d), confirms the presence of CEB in MnSe. Apart from 100 K, surprisingly no significant CEB phenomenon has been observed in other temperature. The exchange bias field $H_\text{CEB}$ obtained under $H_\text{FC}$ = 10 kOe is $\sim$ 735 Oe.

\subsubsection{Magnetocaloric effect}
The change in magnetic entropy is a key parameter for studying the magnetocaloric effect (MCE) and can also serve as an indicator of magnetic phase transitions. The $\Delta$S$_\text{M}$ curve is shown in Fig. S3(b). A pronounced peak is observed in the $\Delta$S$_\text{M}$ curves with a $\Delta$S$_\text{M}$ value of approximately 44 $\times$ 10$^{-4}$ J/kg·K at $H = 50$ kOe around the Néel temperature. 
Additionally, a clear transition from the inverse magnetocaloric effect to the conventional magnetocaloric effect is observed around 220 K, just after the $T_\text{N}$, in good agreement with the $H_\text{C}$ vs $T$ plot [as shown in Fig. S2(d)]. 

\subsection{Raman study} 
\begin{figure*}[t]
\centering
\includegraphics[width=1.85\columnwidth]{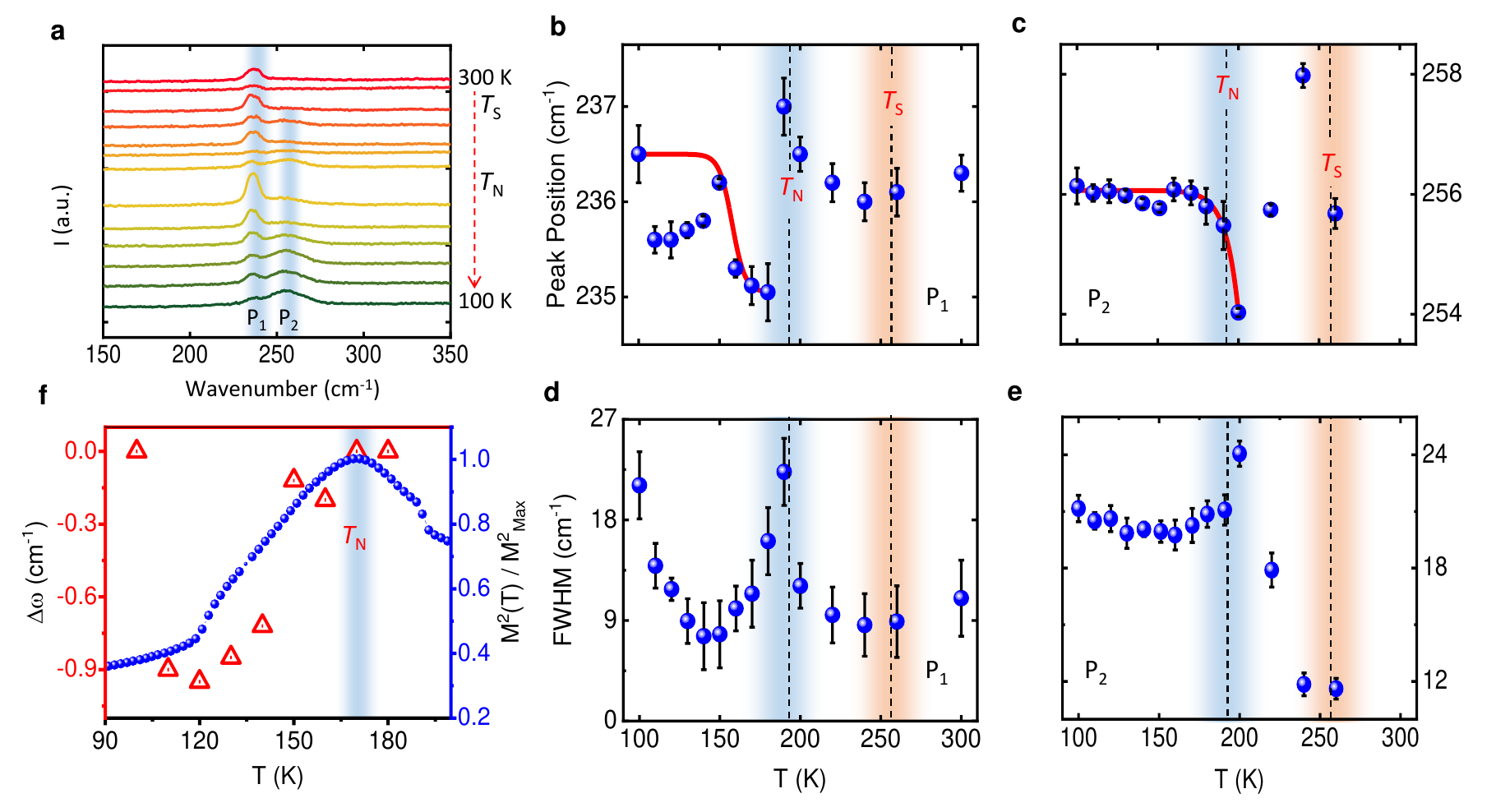}
\caption{\textbf{Spin-Phonon coupling :} (a) Temperature-dependent Raman spectra were recorded across the Néel temperature and structural phase transition, from 300 K to 100 K, in the 150 – 350 cm$^{-1}$ range, capturing the P$_1$ and P$_2$ modes. The variations in peak positions and line widths (FWHM) of P$_1$ and P$_2$ are presented in panels (b–e). P$_3$ is not shown here due to its relatively low intensity compared to the other prominent peaks. The red solid lines in (b) and (c) are fit to the Eq.~\ref{eq_sig}. (f) The deviation from anharmonicity can be correlated to the
abrupt jump of magnetization at around 170 K ($T_\text{N}$). The red triangles denote the $\Delta$$\omega$ values obtained from the shift in P$_1$ mode, while the blue solid spheres correspond to the normalized $M(T)^2$ values obtained from magnetization measurements at $H = 0.1$ kOe (parallel to [111]). Notably, $\Delta\omega$ exhibits a trend closely matching that of the normalized $M(T)^2$.}
\label{f_sp}
\end{figure*}


\subsubsection{Signature of phase transitions}
Raman spectroscopy serves as a powerful tool for investigating phase transitions and various elementary excitations, including lattice vibrations, electron-phonon interactions, and spin-phonon interactions, particularly in reduced dimensions \cite{Du,Bera2025}. To explore the antiferromagnetic (AFM) properties of MnSe and establish the correlation between spin dynamics and phonon modes, we conducted temperature-dependent Raman spectroscopy measurements. Systematic Raman spectra were collected over the temperature range of 300–100 K and analyzed using a Lorentzian multi-function fitting approach. Figure \ref{f_sp}(a) presents representative Raman spectra within the spectral range of 150 cm$^{-1}$ < $\omega$ < 350 cm$^{-1}$. 

 The temperature dependence of peak 1 is plotted in Fig.~\ref{3D}(a,b). The Raman spectrum at 300 K [see Fig.~\ref{3D}(a)] reveals a single distinct asymmetric peak, labeled P$_1$. The observed asymmetry may result from coupling between the phonon and other elementary excitations \cite{PhysRevB.12.4328}.  The position of the peak with temperature exhibits distinct behavior in two regions: before and after the magnetic transition. Upon closer inspection, this peak remains unchanged down to 100 K. However, below 260 K, the asymmetry in P$_1$ causes it to split, resulting in the emergence of a second peak at 256.24 cm$^{-1}$ (P$_2$) [see Fig.~ \ref{3D}(a)]. This newly observed peak is attributed to the transverse optical + longitudinal acoustic (TO + LA) mode, which originates from lattice vibrations \cite{milutinovic2004,PhysRevB.73.155203}.  Beyond 260 K, P$_1$ remains the dominant peak, while P$_2$ gradually intensifies as the temperature decreases, reaching parity with P$_1$ at 170 K. This temperature point is very much near to $T_\text{N}$. Below 170 K, P$_2$ begins to dominate over P$_1$ as the temperature continues to decrease, up to the lowest measured temperature [see Fig.~\ref{3D}(b)]. 

To further explore the temperature-dependent phonon behavior, the corresponding peak frequency position ($\omega$), and line width (FWHM) are extracted from the respective Lorentzian fits to the individual phonon modes. Figs.~\ref{f_sp}(b-e) depict the temperature dependence of the selected phonon energies (of P$_1$ and P$_2$) along with their full width at half maximum (FWHM). For P$_1$, two distinct minima are clearly observed around 170 K and 250 K, while for P$_2$, an anomaly appears at 170 K. These observations suggest phase transitions consistent with those identified in the magnetic response data discussed in Section B.

The variation in the classified peaks at $T$ = 100 K, compared to the room temperature data [Fig. \ref{f_sp}(a)] is readily apparent to the naked eye and may indicate contributions from phonon anharmonicity and spin-phonon coupling. 

\subsubsection{Spin-Phonon (SP) coupling}
In the previous section, we have discussed the signatures of phase transitions in MnSe through phonon characteristics, where peak 2 is identified as a Raman mode sensitive to structural transitions. This peak 2 is located very close to peak 1, with near overlap observed in the mid-temperature range. P$_1$, which persists throughout the entire temperature range, shows a temperature-dependent variation of peak positions that can be fitted below the magnetic transition using the Boltzmann-Sigmoidal equation \cite{Casto2015},
 
\begin{equation}
    \omega(T) = \omega_1 + \frac{\omega^\prime - \omega_1}{1+exp\frac{T-T_0}{dT}}
    \label{eq_sig}
\end{equation}
Here, $\omega^\prime$ and $\omega_1$ denote the upper and lower limits of the sigmoidal curve, respectively. $T_0$ represents the midpoint, while ${dT}$ determines the curve's width. The red solid line in Fig.~\ref{f_sp}(b) represents the fitted result for P$_1$ using Eq.~\ref{eq_sig}. 

However, a closer inspection of the temperature-dependent peak position within the small temperature range of 100~K to 160~K reveals deviations from the Boltzmann-Sigmoidal behavior. Notably, the antiferromagnetic transition occurs at 170~K. Therefore, this deviation from anharmonicity is attributed to spin-phonon coupling \cite{Ghosh2021}. 

In a phenomenological description, weak spin-phonon coupling in the lattice can be described as \cite{Lee}:
\begin{equation}
    \Delta\omega = \lambda_{sp}<S_iS_{i+1}> = \lambda_{sp}S^2\phi(T) 
    \label{eq_sp}
\end{equation}
Here, $<S_iS_{i+1}>$ represents the spin correlation function for spin-ordered phases, capturing the average interaction between adjacent spins in opposite directions, and vanishes in paramagnetic phases where spin interactions are absent. $\Delta\omega$ denotes the deviation from the Boltzmann-Sigmoidal behavior of the peak position, while $\lambda_\text{sp}$ is the spin-phonon (SP) coupling coefficient. In general, $\lambda_\text{sp}$ is related to the magnetic exchange interaction and can be positive or negative, depending on whether the phonon hardens (typically in AFM systems \cite{PhysRevB.75.104118}) or softens (usually in FM systems \cite{Iliev}). The order parameter, $\phi(T)$, is defined as $\phi(T) = 1 - (T/T_\text{N})^\gamma$ with $S = 2$ (Mn$^{+2}$), meaning that $\Delta\omega(T)$ effectively quantifies the strength of SP coupling. For the P$1$ mode, $\lambda_\text{sp}$ is found to be approximately 0.5375 cm$^{-1}$, providing strong evidence of significant spin-phonon interactions driven by the PM-to-AFM transition. This suggests that atoms with stronger bonds play a crucial role in magnetic ordering, particularly in collinear spin states. The obtained $\lambda_\text{sp}$ value aligns with those typically observed in 2D magnets \cite{Ghorai2024}, yet remains lower than those reported for magnetic nanoparticles and bulk antiferromagnetic materials such as Cr$_2$O$_3$, FeF$_2$, and MnF$_2$ \cite{Wu2012, Lockwood1983}.  To extract the exponent $\gamma$ of the order parameter, the normalized deviation ($\Delta\omega$) is fitted as a function of the reduced temperature ($T/T_\text{N}$) using Eq.~\ref{eq_sp}, as presented in Fig. S5(a). The fitting yields a value of $\gamma$ = 5.46.

Using the mean-field approximation, Granado et al. proposed a mechanism to describe phonon renormalization induced by spin-phonon coupling \cite{GranadoE}. In other words $\Delta\omega(T)$ [as expressed in Eq.~\ref{eq_sp}] can be related to magnetization as $\Delta\omega(T) \propto \frac{M^2(T)}{M^2_{max}}$. Here, $M(T)$ represents the average magnetization per magnetic ion at temperature $T$. While this model was originally developed for perovskite structures, it is also highly applicable to 2D magnetic materials \cite{Ghosh2021}. The temperature dependence of $\Delta\omega$ is plotted together with $\frac{M^2(T)}{M^2_\text{max}}$ in Fig.~\ref{f_sp}(f), where both the $\Delta\omega(T)$ data from Raman scattering for peak 1 and the $\frac{M^2(T)}{M^2_\text{max}}$ data from susceptibility measurements display a similar trend [see Fig.~\ref{f_sp}(f)]. The result confirms a linear correlation between the renormalization of phonon positions and the square of the normalized magnetization, in accordance with the proposed model.


The temperature dependence of P$_2$ and P$_3$ is well described by a fit to the previously mentioned Boltzmann-sigmoidal equation, illustrating that each phonon mode exhibits a unique and independent factor $\lambda_\text{sp}$. However, in contrast to P$_1$, the negligible $\Delta\omega$ observed in P$_2$ [see Fig.~\ref{f_sp}(c)] and P$_3$ [not shown] indicates that these modes do not contribute significantly to spin-phonon coupling in the pristine compound. A detailed summary of both P$_1$ and P$_2$ phonon modes is provided in Table~\ref{SP}.

\begin{table}[h!]
	\caption{Details of P$_1$ and P$_2$ modes} 
	\centering 
	\begin{tabular}{|p{0.9cm}|p{1.3cm}|p{1.2cm}|p{0.7cm}|p{1.8cm}|}
		\hline
		Mode  & Position (cm$^{-1})$ & $\lambda_\text{sp}$ (cm$^{-1}$) & $\gamma$ & Importance (in)\\
		\hline
		P$_1$ & 233  & 0.54 & 5.46 & Spin-Phonon coupling \\
		P$_2$ & 256  & 0.21 & 3.38 & Structural-phase transition\\
				\hline
	\end{tabular}
    \label{SP}
\end{table}

\section{Outlook and Conclusion}
In summary, we have presented a comprehensive study of the magnetic and phonon characteristics of single crystalline MnSe, a magnetic semiconductor synthesized using the CVT method. The SC-XRD finds the pristine crystal adopts a Rock Salt-type structure at room temperature. Magnetic measurements [$M(T,H)$] indicate that the material predominantly exhibits antiferromagnetic (AFM) behavior, with the ordering temperature ($T_\text{N}$) of 170 K, alongside a weak ferromagnetic (FM) component. The competing AFM and FM interactions give rise to both spontaneous and conventional exchange bias effects at low temperatures, reflecting the presence of intrinsic phase inhomogeneity. Magnetocaloric and transport data also identify this magnetic transition. 
We employed the temperature-dependent unpolarised Raman scattering study to uncover any possible spin and phonon interaction around $T_\text{N}$ in MnSe. The magnetic transition is well reflected in the observed Raman frequency and FWHM in different Raman modes
from the temperature-dependent Raman measurements. The anomalies below the $T_\text{N}$ suggest the significant spin-phonon interaction parameter with
($\lambda_\text{sp}$) 0.5375 cm$^{-1}$, primarily arising from the Raman mode at 233 cm$^{-1}$. In contrast, the Raman mode at approximately 256 cm$^{-1}$ exhibits sensitivity to a structural phase transition occurring around 250 K. Therefore, these results provide a comprehensive understanding of the structural, magnetic, and vibrational properties of MnSe, highlighting the crucial role of spin-phonon coupling in driving magnetic fluctuations and establishing long-range antiferromagnetic order. Consequently, MnSe emerges as a promising candidate for two-dimensional magnetic spintronic device applications.


\textbf{Acknowledgement:} This work is supported by the National Key R\&D Program of China (grant no. 2022YFA1203902), the National Natural Science Foundation of China (NSFC) (grant nos. 12374108 and 12104052), and the Guangdong Provincial Quantum Science Strategic Initiative (Grant No. GDZX2401002), the Fundamental Research Funds for the Central Universities, the State Key Lab of Luminescent Materials and Devices, South China University of Technology, and GBRCE for Functional Molecular Engineering. 



\appendix
\renewcommand{\thefigure}{A\arabic{figure}}
\renewcommand{\thesection}{A\arabic{section}}
\setcounter{figure}{0}
\setcounter{section}{0}

%
\section{Temperature-dependent Raman spectra}
\begin{figure}
\centering
\includegraphics[width=1\columnwidth]{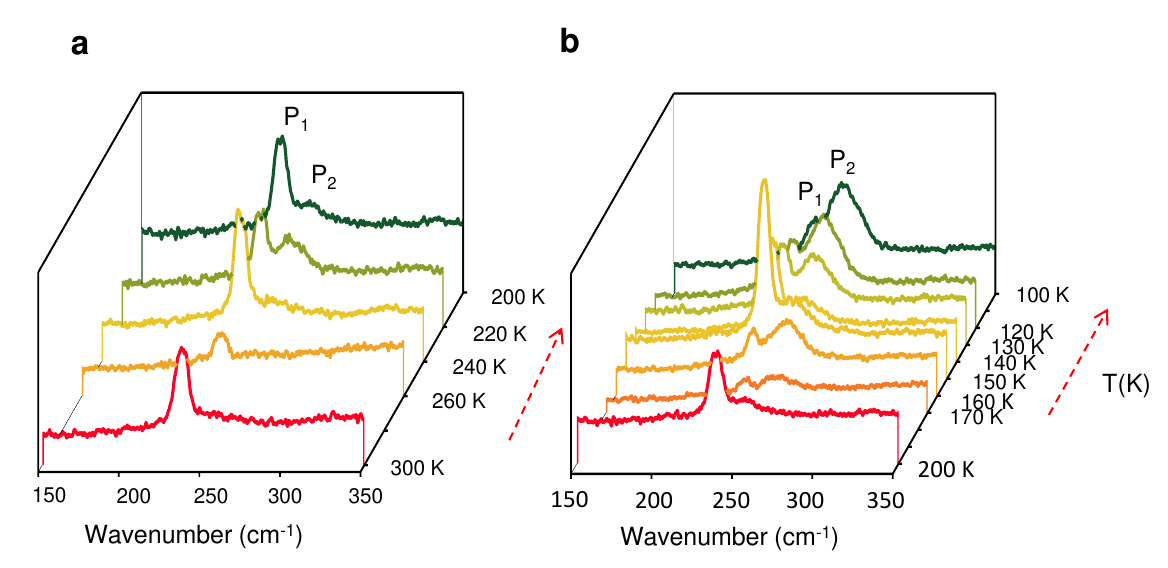}
\centering
\caption{\textbf{3D plot of temperature dependence of unpolarised Raman spectra :} Panels (a,b) display the temperature evolution of P$_1$ and P$_2$ within the 150 to 350 cm$^{-1}$. 
The red dotted arrow indicates the decrease in temperature.}
\vspace{-0.45cm}
\label{3D}
\end{figure}

Figs.~\ref{3D}(a,b) illustrate the temperature evolution of selected Raman modes, P$_1$ and P$_2$, in the range of 300 K to 100 K.



\begin{thebibliography}{45}%
\makeatletter
\providecommand \@ifxundefined [1]{%
 \@ifx{#1\undefined}
}%
\providecommand \@ifnum [1]{%
 \ifnum #1\expandafter \@firstoftwo
 \else \expandafter \@secondoftwo
 \fi
}%
\providecommand \@ifx [1]{%
 \ifx #1\expandafter \@firstoftwo
 \else \expandafter \@secondoftwo
 \fi
}%
\providecommand \natexlab [1]{#1}%
\providecommand \enquote  [1]{``#1''}%
\providecommand \bibnamefont  [1]{#1}%
\providecommand \bibfnamefont [1]{#1}%
\providecommand \citenamefont [1]{#1}%
\providecommand \href@noop [0]{\@secondoftwo}%
\providecommand \href [0]{\begingroup \@sanitize@url \@href}%
\providecommand \@href[1]{\@@startlink{#1}\@@href}%
\providecommand \@@href[1]{\endgroup#1\@@endlink}%
\providecommand \@sanitize@url [0]{\catcode `\\12\catcode `\$12\catcode `\&12\catcode `\#12\catcode `\^12\catcode `\_12\catcode `\%12\relax}%
\providecommand \@@startlink[1]{}%
\providecommand \@@endlink[0]{}%
\providecommand \url  [0]{\begingroup\@sanitize@url \@url }%
\providecommand \@url [1]{\endgroup\@href {#1}{\urlprefix }}%
\providecommand \urlprefix  [0]{URL }%
\providecommand \Eprint [0]{\href }%
\providecommand \doibase [0]{https://doi.org/}%
\providecommand \selectlanguage [0]{\@gobble}%
\providecommand \bibinfo  [0]{\@secondoftwo}%
\providecommand \bibfield  [0]{\@secondoftwo}%
\providecommand \translation [1]{[#1]}%
\providecommand \BibitemOpen [0]{}%
\providecommand \bibitemStop [0]{}%
\providecommand \bibitemNoStop [0]{.\EOS\space}%
\providecommand \EOS [0]{\spacefactor3000\relax}%
\providecommand \BibitemShut  [1]{\csname bibitem#1\endcsname}%
\let\auto@bib@innerbib\@empty
\bibitem [{\citenamefont {Haas}(1970)}]{Haas}%
  \BibitemOpen
  \bibfield  {author} {\bibinfo {author} {\bibfnamefont {C.}~\bibnamefont {Haas}},\ }\bibfield  {title} {\bibinfo {title} {Magnetic semiconductors},\ }\href {https://doi.org/10.1080/10408437008243418} {\bibfield  {journal} {\bibinfo  {journal} {C R C Critical Reviews in Solid State Sciences}\ }\textbf {\bibinfo {volume} {1}},\ \bibinfo {pages} {47} (\bibinfo {year} {1970})},\ \Eprint {https://arxiv.org/abs/https://doi.org/10.1080/10408437008243418} {https://doi.org/10.1080/10408437008243418} \BibitemShut {NoStop}%
\bibitem [{\citenamefont {Jungwirth}\ \emph {et~al.}(2002)\citenamefont {Jungwirth}, \citenamefont {Niu},\ and\ \citenamefont {MacDonald}}]{PhysRevLett.88.207208}%
  \BibitemOpen
  \bibfield  {author} {\bibinfo {author} {\bibfnamefont {T.}~\bibnamefont {Jungwirth}}, \bibinfo {author} {\bibfnamefont {Q.}~\bibnamefont {Niu}},\ and\ \bibinfo {author} {\bibfnamefont {A.~H.}\ \bibnamefont {MacDonald}},\ }\bibfield  {title} {\bibinfo {title} {Anomalous hall effect in ferromagnetic semiconductors},\ }\href {https://doi.org/10.1103/PhysRevLett.88.207208} {\bibfield  {journal} {\bibinfo  {journal} {Phys. Rev. Lett.}\ }\textbf {\bibinfo {volume} {88}},\ \bibinfo {pages} {207208} (\bibinfo {year} {2002})}\BibitemShut {NoStop}%
\bibitem [{\citenamefont {Wilson}\ \emph {et~al.}(2021)\citenamefont {Wilson}, \citenamefont {Lee}, \citenamefont {Cenker}, \citenamefont {Xie}, \citenamefont {Dismukes}, \citenamefont {Telford}, \citenamefont {Fonseca}, \citenamefont {Sivakumar}, \citenamefont {Dean}, \citenamefont {Cao}, \citenamefont {Roy}, \citenamefont {Xu},\ and\ \citenamefont {Zhu}}]{Wilson2021}%
  \BibitemOpen
  \bibfield  {author} {\bibinfo {author} {\bibfnamefont {N.~P.}\ \bibnamefont {Wilson}}, \bibinfo {author} {\bibfnamefont {K.}~\bibnamefont {Lee}}, \bibinfo {author} {\bibfnamefont {J.}~\bibnamefont {Cenker}}, \bibinfo {author} {\bibfnamefont {K.}~\bibnamefont {Xie}}, \bibinfo {author} {\bibfnamefont {A.~H.}\ \bibnamefont {Dismukes}}, \bibinfo {author} {\bibfnamefont {E.~J.}\ \bibnamefont {Telford}}, \bibinfo {author} {\bibfnamefont {J.}~\bibnamefont {Fonseca}}, \bibinfo {author} {\bibfnamefont {S.}~\bibnamefont {Sivakumar}}, \bibinfo {author} {\bibfnamefont {C.}~\bibnamefont {Dean}}, \bibinfo {author} {\bibfnamefont {T.}~\bibnamefont {Cao}}, \bibinfo {author} {\bibfnamefont {X.}~\bibnamefont {Roy}}, \bibinfo {author} {\bibfnamefont {X.}~\bibnamefont {Xu}},\ and\ \bibinfo {author} {\bibfnamefont {X.}~\bibnamefont {Zhu}},\ }\bibfield  {title} {\bibinfo {title} {Interlayer electronic coupling on demand in a 2d magnetic semiconductor},\ }\href {https://doi.org/10.1038/s41563-021-01070-8} {\bibfield  {journal}
  {\bibinfo  {journal} {Nature Materials}\ }\textbf {\bibinfo {volume} {20}},\ \bibinfo {pages} {1657} (\bibinfo {year} {2021})}\BibitemShut {NoStop}%
\bibitem [{\citenamefont {Wolf}\ \emph {et~al.}(2001)\citenamefont {Wolf}, \citenamefont {Awschalom}, \citenamefont {Buhrman}, \citenamefont {Daughton}, \citenamefont {von Molnár}, \citenamefont {Roukes}, \citenamefont {Chtchelkanova},\ and\ \citenamefont {Treger}}]{Wolf2001}%
  \BibitemOpen
  \bibfield  {author} {\bibinfo {author} {\bibfnamefont {S.~A.}\ \bibnamefont {Wolf}}, \bibinfo {author} {\bibfnamefont {D.~D.}\ \bibnamefont {Awschalom}}, \bibinfo {author} {\bibfnamefont {R.~A.}\ \bibnamefont {Buhrman}}, \bibinfo {author} {\bibfnamefont {J.~M.}\ \bibnamefont {Daughton}}, \bibinfo {author} {\bibfnamefont {S.}~\bibnamefont {von Molnár}}, \bibinfo {author} {\bibfnamefont {M.~L.}\ \bibnamefont {Roukes}}, \bibinfo {author} {\bibfnamefont {A.~Y.}\ \bibnamefont {Chtchelkanova}},\ and\ \bibinfo {author} {\bibfnamefont {D.~M.}\ \bibnamefont {Treger}},\ }\bibfield  {title} {\bibinfo {title} {Spintronics: A spin-based electronics vision for the future},\ }\href {https://doi.org/10.1126/science.1065389} {\bibfield  {journal} {\bibinfo  {journal} {Science}\ }\textbf {\bibinfo {volume} {294}},\ \bibinfo {pages} {1488} (\bibinfo {year} {2001})},\ \Eprint {https://arxiv.org/abs/https://www.science.org/doi/pdf/10.1126/science.1065389} {https://www.science.org/doi/pdf/10.1126/science.1065389} \BibitemShut
  {NoStop}%
\bibitem [{\citenamefont {Awschalom}\ and\ \citenamefont {Flatté}(2007)}]{Awschalom}%
  \BibitemOpen
  \bibfield  {author} {\bibinfo {author} {\bibfnamefont {D.~D.}\ \bibnamefont {Awschalom}}\ and\ \bibinfo {author} {\bibfnamefont {M.~E.}\ \bibnamefont {Flatté}},\ }\bibfield  {title} {\bibinfo {title} {Challenges for semiconductor spintronics},\ }\href {https://doi.org/10.1038/nphys551} {\bibfield  {journal} {\bibinfo  {journal} {Nature Physics}\ }\textbf {\bibinfo {volume} {3}},\ \bibinfo {pages} {153} (\bibinfo {year} {2007})}\BibitemShut {NoStop}%
\bibitem [{\citenamefont {Schlesinger}(1998)}]{Schlesinger1998}%
  \BibitemOpen
  \bibfield  {author} {\bibinfo {author} {\bibfnamefont {M.~E.}\ \bibnamefont {Schlesinger}},\ }\bibfield  {title} {\bibinfo {title} {The mn-se (manganese-selenium) system},\ }\href {https://doi.org/10.1361/105497198770341798} {\bibfield  {journal} {\bibinfo  {journal} {Journal of Phase Equilibria}\ }\textbf {\bibinfo {volume} {19}},\ \bibinfo {pages} {588} (\bibinfo {year} {1998})}\BibitemShut {NoStop}%
\bibitem [{\citenamefont {Giebul/towicz}\ \emph {et~al.}(1992)\citenamefont {Giebul/towicz}, \citenamefont {Samarth}, \citenamefont {Luo}, \citenamefont {Furdyna}, \citenamefont {Kl/osowski},\ and\ \citenamefont {Rhyne}}]{PhysRevB.46.12076}%
  \BibitemOpen
  \bibfield  {author} {\bibinfo {author} {\bibfnamefont {T.~M.}\ \bibnamefont {Giebul/towicz}}, \bibinfo {author} {\bibfnamefont {N.}~\bibnamefont {Samarth}}, \bibinfo {author} {\bibfnamefont {H.}~\bibnamefont {Luo}}, \bibinfo {author} {\bibfnamefont {J.~K.}\ \bibnamefont {Furdyna}}, \bibinfo {author} {\bibfnamefont {P.}~\bibnamefont {Kl/osowski}},\ and\ \bibinfo {author} {\bibfnamefont {J.~J.}\ \bibnamefont {Rhyne}},\ }\bibfield  {title} {\bibinfo {title} {Strain-engineered incommensurability in epitaxial heisenberg antiferromagnets},\ }\href {https://doi.org/10.1103/PhysRevB.46.12076} {\bibfield  {journal} {\bibinfo  {journal} {Phys. Rev. B}\ }\textbf {\bibinfo {volume} {46}},\ \bibinfo {pages} {12076} (\bibinfo {year} {1992})}\BibitemShut {NoStop}%
\bibitem [{\citenamefont {Rao}\ and\ \citenamefont {Pisharody}(1976)}]{RAO1976207}%
  \BibitemOpen
  \bibfield  {author} {\bibinfo {author} {\bibfnamefont {C.}~\bibnamefont {Rao}}\ and\ \bibinfo {author} {\bibfnamefont {K.}~\bibnamefont {Pisharody}},\ }\bibfield  {title} {\bibinfo {title} {Transition metal sulfides},\ }\href {https://doi.org/https://doi.org/10.1016/0079-6786(76)90009-1} {\bibfield  {journal} {\bibinfo  {journal} {Progress in Solid State Chemistry}\ }\textbf {\bibinfo {volume} {10}},\ \bibinfo {pages} {207} (\bibinfo {year} {1976})}\BibitemShut {NoStop}%
\bibitem [{\citenamefont {Efrem~D'Sa}\ \emph {et~al.}(2004)\citenamefont {Efrem~D'Sa}, \citenamefont {Bhobe}, \citenamefont {Priolkar}, \citenamefont {Das}, \citenamefont {Krishna}, \citenamefont {Sarode},\ and\ \citenamefont {Prabhu}}]{EfremDSa2004}%
  \BibitemOpen
  \bibfield  {author} {\bibinfo {author} {\bibfnamefont {J.~B.~C.}\ \bibnamefont {Efrem~D'Sa}}, \bibinfo {author} {\bibfnamefont {P.~A.}\ \bibnamefont {Bhobe}}, \bibinfo {author} {\bibfnamefont {K.~R.}\ \bibnamefont {Priolkar}}, \bibinfo {author} {\bibfnamefont {A.}~\bibnamefont {Das}}, \bibinfo {author} {\bibfnamefont {P.~S.~R.}\ \bibnamefont {Krishna}}, \bibinfo {author} {\bibfnamefont {P.~R.}\ \bibnamefont {Sarode}},\ and\ \bibinfo {author} {\bibfnamefont {R.~B.}\ \bibnamefont {Prabhu}},\ }\bibfield  {title} {\bibinfo {title} {Low temperature magnetic structure of mnse},\ }\href {https://doi.org/10.1007/BF02704977} {\bibfield  {journal} {\bibinfo  {journal} {Pramana}\ }\textbf {\bibinfo {volume} {63}},\ \bibinfo {pages} {227} (\bibinfo {year} {2004})}\BibitemShut {NoStop}%
\bibitem [{\citenamefont {Huang}\ \emph {et~al.}(2019)\citenamefont {Huang}, \citenamefont {Wang}, \citenamefont {Chang}, \citenamefont {Lee}, \citenamefont {Huang}, \citenamefont {Wang},\ and\ \citenamefont {Wu}}]{Huang2019}%
  \BibitemOpen
  \bibfield  {author} {\bibinfo {author} {\bibfnamefont {C.-H.}\ \bibnamefont {Huang}}, \bibinfo {author} {\bibfnamefont {C.-W.}\ \bibnamefont {Wang}}, \bibinfo {author} {\bibfnamefont {C.-C.}\ \bibnamefont {Chang}}, \bibinfo {author} {\bibfnamefont {Y.-C.}\ \bibnamefont {Lee}}, \bibinfo {author} {\bibfnamefont {G.-T.}\ \bibnamefont {Huang}}, \bibinfo {author} {\bibfnamefont {M.-J.}\ \bibnamefont {Wang}},\ and\ \bibinfo {author} {\bibfnamefont {M.-K.}\ \bibnamefont {Wu}},\ }\bibfield  {title} {\bibinfo {title} {Anomalous magnetic properties in mn(se, s) system},\ }\href {https://doi.org/https://doi.org/10.1016/j.jmmm.2019.03.105} {\bibfield  {journal} {\bibinfo  {journal} {Journal of Magnetism and Magnetic Materials}\ }\textbf {\bibinfo {volume} {483}},\ \bibinfo {pages} {205} (\bibinfo {year} {2019})}\BibitemShut {NoStop}%
\bibitem [{\citenamefont {Sines}\ \emph {et~al.}(2010)\citenamefont {Sines}, \citenamefont {Misra}, \citenamefont {Schiffer},\ and\ \citenamefont {Schaak}}]{https://doi.org/10.1002/anie.201001213}%
  \BibitemOpen
  \bibfield  {author} {\bibinfo {author} {\bibfnamefont {I.}~\bibnamefont {Sines}}, \bibinfo {author} {\bibfnamefont {R.}~\bibnamefont {Misra}}, \bibinfo {author} {\bibfnamefont {P.}~\bibnamefont {Schiffer}},\ and\ \bibinfo {author} {\bibfnamefont {R.}~\bibnamefont {Schaak}},\ }\bibfield  {title} {\bibinfo {title} {Colloidal synthesis of non-equilibrium wurtzite-type mnse},\ }\href {https://doi.org/https://doi.org/10.1002/anie.201001213} {\bibfield  {journal} {\bibinfo  {journal} {Angewandte Chemie International Edition}\ }\textbf {\bibinfo {volume} {49}},\ \bibinfo {pages} {4638} (\bibinfo {year} {2010})},\ \Eprint {https://arxiv.org/abs/https://onlinelibrary.wiley.com/doi/pdf/10.1002/anie.201001213} {https://onlinelibrary.wiley.com/doi/pdf/10.1002/anie.201001213} \BibitemShut {NoStop}%
\bibitem [{\citenamefont {Squire}(9 11)}]{Squire1939}%
  \BibitemOpen
  \bibfield  {author} {\bibinfo {author} {\bibfnamefont {C.~F.}\ \bibnamefont {Squire}},\ }\bibfield  {title} {\bibinfo {title} {Antiferromagnetism in some manganous compounds},\ }\href {https://doi.org/10.1103/PhysRev.56.922} {\bibfield  {journal} {\bibinfo  {journal} {Phys. Rev.}\ }\textbf {\bibinfo {volume} {56}},\ \bibinfo {pages} {922} (\bibinfo {year} {1939-11})}\BibitemShut {NoStop}%
\bibitem [{\citenamefont {Lindsay}(1 11)}]{Lindsay1951}%
  \BibitemOpen
  \bibfield  {author} {\bibinfo {author} {\bibfnamefont {R.}~\bibnamefont {Lindsay}},\ }\bibfield  {title} {\bibinfo {title} {Magnetic susceptibility of manganese selenide},\ }\href {https://doi.org/10.1103/PhysRev.84.569} {\bibfield  {journal} {\bibinfo  {journal} {Phys. Rev.}\ }\textbf {\bibinfo {volume} {84}},\ \bibinfo {pages} {569} (\bibinfo {year} {1951-11})}\BibitemShut {NoStop}%
\bibitem [{\citenamefont {Ito}\ \emph {et~al.}(8 02)\citenamefont {Ito}, \citenamefont {Ito},\ and\ \citenamefont {Oka}}]{Ito1978}%
  \BibitemOpen
  \bibfield  {author} {\bibinfo {author} {\bibfnamefont {T.}~\bibnamefont {Ito}}, \bibinfo {author} {\bibfnamefont {K.}~\bibnamefont {Ito}},\ and\ \bibinfo {author} {\bibfnamefont {M.}~\bibnamefont {Oka}},\ }\bibfield  {title} {\bibinfo {title} {Magnetic susceptibility, thermal expansion and electrical resistivity of mnse},\ }\href {https://doi.org/10.1143/JJAP.17.371} {\bibfield  {journal} {\bibinfo  {journal} {Japanese Journal of Applied Physics}\ }\textbf {\bibinfo {volume} {17}},\ \bibinfo {pages} {371} (\bibinfo {year} {1978-02})}\BibitemShut {NoStop}%
\bibitem [{\citenamefont {Popovi\ifmmode~\acute{c}\else \'{c}\fi{}}\ and\ \citenamefont {Milutinovi\ifmmode~\acute{c}\else \'{c}\fi{}}(6 04)}]{Popoviifmmodeacutecelsecfi2006}%
  \BibitemOpen
  \bibfield  {author} {\bibinfo {author} {\bibfnamefont {Z.~V.}\ \bibnamefont {Popovi\ifmmode~\acute{c}\else \'{c}\fi{}}}\ and\ \bibinfo {author} {\bibfnamefont {A.}~\bibnamefont {Milutinovi\ifmmode~\acute{c}\else \'{c}\fi{}}},\ }\bibfield  {title} {\bibinfo {title} {Far-infrared reflectivity and raman scattering study of $\ensuremath{\alpha}\text{\ensuremath{-}}\mathrm{Mn}\mathrm{Se}$},\ }\href {https://doi.org/10.1103/PhysRevB.73.155203} {\bibfield  {journal} {\bibinfo  {journal} {Phys. Rev. B}\ }\textbf {\bibinfo {volume} {73}},\ \bibinfo {pages} {155203} (\bibinfo {year} {2006-04})}\BibitemShut {NoStop}%
\bibitem [{\citenamefont {O’Hara}\ \emph {et~al.}(2018)\citenamefont {O’Hara}, \citenamefont {Zhu}, \citenamefont {Trout}, \citenamefont {Ahmed}, \citenamefont {Luo}, \citenamefont {Lee}, \citenamefont {Brenner}, \citenamefont {Rajan}, \citenamefont {Gupta}, \citenamefont {McComb},\ and\ \citenamefont {Kawakami}}]{O’Hara2018}%
  \BibitemOpen
  \bibfield  {author} {\bibinfo {author} {\bibfnamefont {D.}~\bibnamefont {O’Hara}}, \bibinfo {author} {\bibfnamefont {T.}~\bibnamefont {Zhu}}, \bibinfo {author} {\bibfnamefont {A.~H.}\ \bibnamefont {Trout}}, \bibinfo {author} {\bibfnamefont {A.~S.}\ \bibnamefont {Ahmed}}, \bibinfo {author} {\bibfnamefont {Y.~K.}\ \bibnamefont {Luo}}, \bibinfo {author} {\bibfnamefont {C.~H.}\ \bibnamefont {Lee}}, \bibinfo {author} {\bibfnamefont {M.~R.}\ \bibnamefont {Brenner}}, \bibinfo {author} {\bibfnamefont {S.}~\bibnamefont {Rajan}}, \bibinfo {author} {\bibfnamefont {J.~A.}\ \bibnamefont {Gupta}}, \bibinfo {author} {\bibfnamefont {D.~W.}\ \bibnamefont {McComb}},\ and\ \bibinfo {author} {\bibfnamefont {R.~K.}\ \bibnamefont {Kawakami}},\ }\bibfield  {title} {\bibinfo {title} {Room temperature intrinsic ferromagnetism in epitaxial manganese selenide films in the monolayer limit},\ }\href {https://doi.org/10.1021/acs.nanolett.8b00683} {\bibfield  {journal} {\bibinfo  {journal} {Nano Lett.}\ }\textbf {\bibinfo {volume}
  {18}},\ \bibinfo {pages} {3125} (\bibinfo {year} {2018})}\BibitemShut {NoStop}%
\bibitem [{\citenamefont {Hung}\ \emph {et~al.}(2021)\citenamefont {Hung}, \citenamefont {Huang}, \citenamefont {Deng}, \citenamefont {Ou}, \citenamefont {Chen}, \citenamefont {Wu}, \citenamefont {Huyan}, \citenamefont {Chu}, \citenamefont {Chen},\ and\ \citenamefont {Lee}}]{Hung2021}%
  \BibitemOpen
  \bibfield  {author} {\bibinfo {author} {\bibfnamefont {T.~L.}\ \bibnamefont {Hung}}, \bibinfo {author} {\bibfnamefont {C.~H.}\ \bibnamefont {Huang}}, \bibinfo {author} {\bibfnamefont {L.~Z.}\ \bibnamefont {Deng}}, \bibinfo {author} {\bibfnamefont {M.~N.}\ \bibnamefont {Ou}}, \bibinfo {author} {\bibfnamefont {Y.~Y.}\ \bibnamefont {Chen}}, \bibinfo {author} {\bibfnamefont {M.~K.}\ \bibnamefont {Wu}}, \bibinfo {author} {\bibfnamefont {S.~Y.}\ \bibnamefont {Huyan}}, \bibinfo {author} {\bibfnamefont {C.~W.}\ \bibnamefont {Chu}}, \bibinfo {author} {\bibfnamefont {P.~J.}\ \bibnamefont {Chen}},\ and\ \bibinfo {author} {\bibfnamefont {T.~K.}\ \bibnamefont {Lee}},\ }\bibfield  {title} {\bibinfo {title} {Pressure induced superconductivity in mnse},\ }\href {https://doi.org/10.1038/s41467-021-25721-1} {\bibfield  {journal} {\bibinfo  {journal} {Nature Communications}\ }\textbf {\bibinfo {volume} {12}},\ \bibinfo {pages} {5436} (\bibinfo {year} {2021})}\BibitemShut {NoStop}%
\bibitem [{\citenamefont {Wang}\ \emph {et~al.}(2016)\citenamefont {Wang}, \citenamefont {Bai}, \citenamefont {Wen}, \citenamefont {Yang}, \citenamefont {Gou}, \citenamefont {Xiao}, \citenamefont {Chow}, \citenamefont {Pravica}, \citenamefont {Yang},\ and\ \citenamefont {Zhao}}]{Wang2016a}%
  \BibitemOpen
  \bibfield  {author} {\bibinfo {author} {\bibfnamefont {Y.}~\bibnamefont {Wang}}, \bibinfo {author} {\bibfnamefont {L.}~\bibnamefont {Bai}}, \bibinfo {author} {\bibfnamefont {T.}~\bibnamefont {Wen}}, \bibinfo {author} {\bibfnamefont {L.}~\bibnamefont {Yang}}, \bibinfo {author} {\bibfnamefont {H.}~\bibnamefont {Gou}}, \bibinfo {author} {\bibfnamefont {Y.}~\bibnamefont {Xiao}}, \bibinfo {author} {\bibfnamefont {P.}~\bibnamefont {Chow}}, \bibinfo {author} {\bibfnamefont {M.}~\bibnamefont {Pravica}}, \bibinfo {author} {\bibfnamefont {W.}~\bibnamefont {Yang}},\ and\ \bibinfo {author} {\bibfnamefont {Y.}~\bibnamefont {Zhao}},\ }\bibfield  {title} {\bibinfo {title} {Giant pressure-driven lattice collapse coupled with intermetallic bonding and spin-state transition in manganese chalcogenides},\ }\href {https://doi.org/https://doi.org/10.1002/anie.201605410} {\bibfield  {journal} {\bibinfo  {journal} {Angewandte Chemie International Edition}\ }\textbf {\bibinfo {volume} {55}},\ \bibinfo {pages} {10350} (\bibinfo
  {year} {2016})},\ \Eprint {https://arxiv.org/abs/https://onlinelibrary.wiley.com/doi/pdf/10.1002/anie.201605410} {https://onlinelibrary.wiley.com/doi/pdf/10.1002/anie.201605410} \BibitemShut {NoStop}%
\bibitem [{\citenamefont {Prasad}\ \emph {et~al.}(1991)\citenamefont {Prasad}, \citenamefont {Pandit}, \citenamefont {Ansari},\ and\ \citenamefont {Singh}}]{Prasad1991}%
  \BibitemOpen
  \bibfield  {author} {\bibinfo {author} {\bibfnamefont {M.}~\bibnamefont {Prasad}}, \bibinfo {author} {\bibfnamefont {A.}~\bibnamefont {Pandit}}, \bibinfo {author} {\bibfnamefont {T.}~\bibnamefont {Ansari}},\ and\ \bibinfo {author} {\bibfnamefont {R.}~\bibnamefont {Singh}},\ }\bibfield  {title} {\bibinfo {title} {Electrical transport properties of manganese selenide},\ }\href {https://doi.org/https://doi.org/10.1016/0254-0584(91)90147-M} {\bibfield  {journal} {\bibinfo  {journal} {Materials Chemistry and Physics}\ }\textbf {\bibinfo {volume} {30}},\ \bibinfo {pages} {13} (\bibinfo {year} {1991})}\BibitemShut {NoStop}%
\bibitem [{\citenamefont {Lei}\ \emph {et~al.}(2006)\citenamefont {Lei}, \citenamefont {Tang},\ and\ \citenamefont {Zheng}}]{Lei2006}%
  \BibitemOpen
  \bibfield  {author} {\bibinfo {author} {\bibfnamefont {S.}~\bibnamefont {Lei}}, \bibinfo {author} {\bibfnamefont {K.}~\bibnamefont {Tang}},\ and\ \bibinfo {author} {\bibfnamefont {H.}~\bibnamefont {Zheng}},\ }\bibfield  {title} {\bibinfo {title} {Solvothermal synthesis of a-mnse uniform nanospheres and nanorods},\ }\href {https://doi.org/https://doi.org/10.1016/j.matlet.2005.11.082} {\bibfield  {journal} {\bibinfo  {journal} {Materials Letters}\ }\textbf {\bibinfo {volume} {60}},\ \bibinfo {pages} {1625} (\bibinfo {year} {2006})}\BibitemShut {NoStop}%
\bibitem [{\citenamefont {Sahoo}\ \emph {et~al.}(2018)\citenamefont {Sahoo}, \citenamefont {Pazhamalai}, \citenamefont {Krishnamoorthy},\ and\ \citenamefont {Kim}}]{Sahoo2018}%
  \BibitemOpen
  \bibfield  {author} {\bibinfo {author} {\bibfnamefont {S.}~\bibnamefont {Sahoo}}, \bibinfo {author} {\bibfnamefont {P.}~\bibnamefont {Pazhamalai}}, \bibinfo {author} {\bibfnamefont {K.}~\bibnamefont {Krishnamoorthy}},\ and\ \bibinfo {author} {\bibfnamefont {S.-J.}\ \bibnamefont {Kim}},\ }\bibfield  {title} {\bibinfo {title} {Hydrothermally prepared a-mnse nanoparticles as a new pseudocapacitive electrode material for supercapacitor},\ }\href {https://doi.org/https://doi.org/10.1016/j.electacta.2018.02.116} {\bibfield  {journal} {\bibinfo  {journal} {Electrochimica Acta}\ }\textbf {\bibinfo {volume} {268}},\ \bibinfo {pages} {403} (\bibinfo {year} {2018})}\BibitemShut {NoStop}%
\bibitem [{\citenamefont {Milutinovi\ifmmode~\acute{c}\else \'{c}\fi{}}\ \emph {et~al.}(2002)\citenamefont {Milutinovi\ifmmode~\acute{c}\else \'{c}\fi{}}, \citenamefont {Tomi\ifmmode~\acute{c}\else \'{c}\fi{}}, \citenamefont {Devi\ifmmode~\acute{c}\else \'{c}\fi{}}, \citenamefont {Milutinovi\ifmmode~\acute{c}\else \'{c}\fi{}},\ and\ \citenamefont {Popovi\ifmmode~\acute{c}\else \'{c}\fi{}}}]{PhysRevB.66.012302}%
  \BibitemOpen
  \bibfield  {author} {\bibinfo {author} {\bibfnamefont {A.}~\bibnamefont {Milutinovi\ifmmode~\acute{c}\else \'{c}\fi{}}}, \bibinfo {author} {\bibfnamefont {N.}~\bibnamefont {Tomi\ifmmode~\acute{c}\else \'{c}\fi{}}}, \bibinfo {author} {\bibfnamefont {S.}~\bibnamefont {Devi\ifmmode~\acute{c}\else \'{c}\fi{}}}, \bibinfo {author} {\bibfnamefont {P.}~\bibnamefont {Milutinovi\ifmmode~\acute{c}\else \'{c}\fi{}}},\ and\ \bibinfo {author} {\bibfnamefont {Z.~V.}\ \bibnamefont {Popovi\ifmmode~\acute{c}\else \'{c}\fi{}}},\ }\bibfield  {title} {\bibinfo {title} {Raman scattering by spin excitations in \ensuremath{\alpha}-mnse},\ }\href {https://doi.org/10.1103/PhysRevB.66.012302} {\bibfield  {journal} {\bibinfo  {journal} {Phys. Rev. B}\ }\textbf {\bibinfo {volume} {66}},\ \bibinfo {pages} {012302} (\bibinfo {year} {2002})}\BibitemShut {NoStop}%
\bibitem [{\citenamefont {Raman}\ \emph {et~al.}(2019)\citenamefont {Raman}, \citenamefont {Chinnadurai}, \citenamefont {Rajmohan}, \citenamefont {Chebrolu}, \citenamefont {Rajangam},\ and\ \citenamefont {Kim}}]{Raman2019}%
  \BibitemOpen
  \bibfield  {author} {\bibinfo {author} {\bibfnamefont {V.}~\bibnamefont {Raman}}, \bibinfo {author} {\bibfnamefont {D.}~\bibnamefont {Chinnadurai}}, \bibinfo {author} {\bibfnamefont {R.}~\bibnamefont {Rajmohan}}, \bibinfo {author} {\bibfnamefont {V.~T.}\ \bibnamefont {Chebrolu}}, \bibinfo {author} {\bibfnamefont {V.}~\bibnamefont {Rajangam}},\ and\ \bibinfo {author} {\bibfnamefont {H.-J.}\ \bibnamefont {Kim}},\ }\bibfield  {title} {\bibinfo {title} {Transition metal chalcogenide based mnse heterostructured with nico2o4 as a new high performance electrode material for capacitive energy storage},\ }\href {https://doi.org/10.1039/C9NJ02711D} {\bibfield  {journal} {\bibinfo  {journal} {New J. Chem.}\ }\textbf {\bibinfo {volume} {43}},\ \bibinfo {pages} {12630} (\bibinfo {year} {2019})}\BibitemShut {NoStop}%
\bibitem [{\citenamefont {Miao}\ \emph {et~al.}(2021)\citenamefont {Miao}, \citenamefont {Fang}, \citenamefont {Zhu}, \citenamefont {Zhou}, \citenamefont {Ye}, \citenamefont {Yan}, \citenamefont {Cao}, \citenamefont {Wang}, \citenamefont {Xu},\ and\ \citenamefont {Xie}}]{Miao2021}%
  \BibitemOpen
  \bibfield  {author} {\bibinfo {author} {\bibfnamefont {C.}~\bibnamefont {Miao}}, \bibinfo {author} {\bibfnamefont {Y.}~\bibnamefont {Fang}}, \bibinfo {author} {\bibfnamefont {K.}~\bibnamefont {Zhu}}, \bibinfo {author} {\bibfnamefont {C.}~\bibnamefont {Zhou}}, \bibinfo {author} {\bibfnamefont {K.}~\bibnamefont {Ye}}, \bibinfo {author} {\bibfnamefont {J.}~\bibnamefont {Yan}}, \bibinfo {author} {\bibfnamefont {D.}~\bibnamefont {Cao}}, \bibinfo {author} {\bibfnamefont {G.}~\bibnamefont {Wang}}, \bibinfo {author} {\bibfnamefont {P.}~\bibnamefont {Xu}},\ and\ \bibinfo {author} {\bibfnamefont {C.}~\bibnamefont {Xie}},\ }\bibfield  {title} {\bibinfo {title} {Binder-free ultrathin a-mnse nanosheets for high performance supercapacitor},\ }\href {https://doi.org/https://doi.org/10.1016/j.jallcom.2021.161004} {\bibfield  {journal} {\bibinfo  {journal} {Journal of Alloys and Compounds}\ }\textbf {\bibinfo {volume} {885}},\ \bibinfo {pages} {161004} (\bibinfo {year} {2021})}\BibitemShut {NoStop}%
\bibitem [{\citenamefont {Zhang}\ \emph {et~al.}(2021{\natexlab{a}})\citenamefont {Zhang}, \citenamefont {Zhao}, \citenamefont {Shen}, \citenamefont {Tao}, \citenamefont {Li}, \citenamefont {Wu}, \citenamefont {Li}, \citenamefont {Yang}, \citenamefont {Li}, \citenamefont {Song}, \citenamefont {Zhang}, \citenamefont {Huang}, \citenamefont {Zhang}, \citenamefont {Zhou}, \citenamefont {Liu},\ and\ \citenamefont {Duan}}]{ZuchengZhang2021}%
  \BibitemOpen
  \bibfield  {author} {\bibinfo {author} {\bibfnamefont {Z.}~\bibnamefont {Zhang}}, \bibinfo {author} {\bibfnamefont {B.}~\bibnamefont {Zhao}}, \bibinfo {author} {\bibfnamefont {D.}~\bibnamefont {Shen}}, \bibinfo {author} {\bibfnamefont {Q.}~\bibnamefont {Tao}}, \bibinfo {author} {\bibfnamefont {B.}~\bibnamefont {Li}}, \bibinfo {author} {\bibfnamefont {R.}~\bibnamefont {Wu}}, \bibinfo {author} {\bibfnamefont {B.}~\bibnamefont {Li}}, \bibinfo {author} {\bibfnamefont {X.}~\bibnamefont {Yang}}, \bibinfo {author} {\bibfnamefont {J.}~\bibnamefont {Li}}, \bibinfo {author} {\bibfnamefont {R.}~\bibnamefont {Song}}, \bibinfo {author} {\bibfnamefont {H.}~\bibnamefont {Zhang}}, \bibinfo {author} {\bibfnamefont {Z.}~\bibnamefont {Huang}}, \bibinfo {author} {\bibfnamefont {Z.}~\bibnamefont {Zhang}}, \bibinfo {author} {\bibfnamefont {J.}~\bibnamefont {Zhou}}, \bibinfo {author} {\bibfnamefont {Y.}~\bibnamefont {Liu}},\ and\ \bibinfo {author} {\bibfnamefont {X.}~\bibnamefont {Duan}},\ }\bibfield  {title} {\bibinfo {title}
  {Synthesis of ultrathin 2d nonlayered a-mnse nanosheets, mnse/ws2 heterojunction for high-performance photodetectors},\ }\href {https://doi.org/https://doi.org/10.1002/sstr.202100028} {\bibfield  {journal} {\bibinfo  {journal} {Small Structures}\ }\textbf {\bibinfo {volume} {2}},\ \bibinfo {pages} {2100028} (\bibinfo {year} {2021}{\natexlab{a}})},\ \Eprint {https://arxiv.org/abs/https://onlinelibrary.wiley.com/doi/pdf/10.1002/sstr.202100028} {https://onlinelibrary.wiley.com/doi/pdf/10.1002/sstr.202100028} \BibitemShut {NoStop}%
\bibitem [{\citenamefont {Park}\ \emph {et~al.}(2022)\citenamefont {Park}, \citenamefont {Lee}, \citenamefont {Park},\ and\ \citenamefont {Kang}}]{Park2022}%
  \BibitemOpen
  \bibfield  {author} {\bibinfo {author} {\bibfnamefont {J.-S.}\ \bibnamefont {Park}}, \bibinfo {author} {\bibfnamefont {A.}~\bibnamefont {Lee}}, \bibinfo {author} {\bibfnamefont {G.~D.}\ \bibnamefont {Park}},\ and\ \bibinfo {author} {\bibfnamefont {Y.~C.}\ \bibnamefont {Kang}},\ }\bibfield  {title} {\bibinfo {title} {Synthesis of mnse@c yolk-shell nanospheres via a water vapor-assisted strategy for use as anode in sodium-ion batteries},\ }\href {https://doi.org/https://doi.org/10.1002/er.7323} {\bibfield  {journal} {\bibinfo  {journal} {International Journal of Energy Research}\ }\textbf {\bibinfo {volume} {46}},\ \bibinfo {pages} {2500} (\bibinfo {year} {2022})},\ \Eprint {https://arxiv.org/abs/https://onlinelibrary.wiley.com/doi/pdf/10.1002/er.7323} {https://onlinelibrary.wiley.com/doi/pdf/10.1002/er.7323} \BibitemShut {NoStop}%
\bibitem [{\citenamefont {Zhang}\ \emph {et~al.}(2021{\natexlab{b}})\citenamefont {Zhang}, \citenamefont {Zhao}, \citenamefont {Shen}, \citenamefont {Tao}, \citenamefont {Li}, \citenamefont {Wu}, \citenamefont {Li}, \citenamefont {Yang}, \citenamefont {Li}, \citenamefont {Song}, \citenamefont {Zhang}, \citenamefont {Huang}, \citenamefont {Zhang}, \citenamefont {Zhou}, \citenamefont {Liu},\ and\ \citenamefont {Duan}}]{Zhang2021}%
  \BibitemOpen
  \bibfield  {author} {\bibinfo {author} {\bibfnamefont {Z.}~\bibnamefont {Zhang}}, \bibinfo {author} {\bibfnamefont {B.}~\bibnamefont {Zhao}}, \bibinfo {author} {\bibfnamefont {D.}~\bibnamefont {Shen}}, \bibinfo {author} {\bibfnamefont {Q.}~\bibnamefont {Tao}}, \bibinfo {author} {\bibfnamefont {B.}~\bibnamefont {Li}}, \bibinfo {author} {\bibfnamefont {R.}~\bibnamefont {Wu}}, \bibinfo {author} {\bibfnamefont {B.}~\bibnamefont {Li}}, \bibinfo {author} {\bibfnamefont {X.}~\bibnamefont {Yang}}, \bibinfo {author} {\bibfnamefont {J.}~\bibnamefont {Li}}, \bibinfo {author} {\bibfnamefont {R.}~\bibnamefont {Song}}, \bibinfo {author} {\bibfnamefont {H.}~\bibnamefont {Zhang}}, \bibinfo {author} {\bibfnamefont {Z.}~\bibnamefont {Huang}}, \bibinfo {author} {\bibfnamefont {Z.}~\bibnamefont {Zhang}}, \bibinfo {author} {\bibfnamefont {J.}~\bibnamefont {Zhou}}, \bibinfo {author} {\bibfnamefont {Y.}~\bibnamefont {Liu}},\ and\ \bibinfo {author} {\bibfnamefont {X.}~\bibnamefont {Duan}},\ }\bibfield  {title} {\bibinfo {title}
  {Synthesis of ultrathin 2d nonlayered a-mnse nanosheets, mnse/ws2 heterojunction for high-performance photodetectors},\ }\href {https://doi.org/https://doi.org/10.1002/sstr.202100028} {\bibfield  {journal} {\bibinfo  {journal} {Small Structures}\ }\textbf {\bibinfo {volume} {2}},\ \bibinfo {pages} {2100028} (\bibinfo {year} {2021}{\natexlab{b}})},\ \Eprint {https://arxiv.org/abs/https://onlinelibrary.wiley.com/doi/pdf/10.1002/sstr.202100028} {https://onlinelibrary.wiley.com/doi/pdf/10.1002/sstr.202100028} \BibitemShut {NoStop}%
\bibitem [{\citenamefont {Zhou}\ \emph {et~al.}(2022)\citenamefont {Zhou}, \citenamefont {Zhang}, \citenamefont {Wang}, \citenamefont {Li}, \citenamefont {Xu}, \citenamefont {Li}, \citenamefont {Ding}, \citenamefont {Liu}, \citenamefont {Li}, \citenamefont {Xie}, \citenamefont {Yang}, \citenamefont {Ma},\ and\ \citenamefont {Zhai}}]{Zhou2022}%
  \BibitemOpen
  \bibfield  {author} {\bibinfo {author} {\bibfnamefont {N.}~\bibnamefont {Zhou}}, \bibinfo {author} {\bibfnamefont {Z.}~\bibnamefont {Zhang}}, \bibinfo {author} {\bibfnamefont {F.}~\bibnamefont {Wang}}, \bibinfo {author} {\bibfnamefont {J.}~\bibnamefont {Li}}, \bibinfo {author} {\bibfnamefont {X.}~\bibnamefont {Xu}}, \bibinfo {author} {\bibfnamefont {H.}~\bibnamefont {Li}}, \bibinfo {author} {\bibfnamefont {S.}~\bibnamefont {Ding}}, \bibinfo {author} {\bibfnamefont {J.}~\bibnamefont {Liu}}, \bibinfo {author} {\bibfnamefont {X.}~\bibnamefont {Li}}, \bibinfo {author} {\bibfnamefont {Y.}~\bibnamefont {Xie}}, \bibinfo {author} {\bibfnamefont {R.}~\bibnamefont {Yang}}, \bibinfo {author} {\bibfnamefont {Y.}~\bibnamefont {Ma}},\ and\ \bibinfo {author} {\bibfnamefont {T.}~\bibnamefont {Zhai}},\ }\bibfield  {title} {\bibinfo {title} {Spin ordering induced broadband photodetection based on two-dimensional magnetic semiconductor a-mnse},\ }\href {https://doi.org/https://doi.org/10.1002/advs.202202177} {\bibfield
  {journal} {\bibinfo  {journal} {Advanced Science}\ }\textbf {\bibinfo {volume} {9}},\ \bibinfo {pages} {2202177} (\bibinfo {year} {2022})},\ \Eprint {https://arxiv.org/abs/https://onlinelibrary.wiley.com/doi/pdf/10.1002/advs.202202177} {https://onlinelibrary.wiley.com/doi/pdf/10.1002/advs.202202177} \BibitemShut {NoStop}%
\bibitem [{\citenamefont {Zhu}\ \emph {et~al.}(2023)\citenamefont {Zhu}, \citenamefont {Xu}, \citenamefont {Tan},\ and\ \citenamefont {Wang}}]{Zhu2023}%
  \BibitemOpen
  \bibfield  {author} {\bibinfo {author} {\bibfnamefont {M.}~\bibnamefont {Zhu}}, \bibinfo {author} {\bibfnamefont {H.}~\bibnamefont {Xu}}, \bibinfo {author} {\bibfnamefont {Z.}~\bibnamefont {Tan}},\ and\ \bibinfo {author} {\bibfnamefont {L.}~\bibnamefont {Wang}},\ }\bibfield  {title} {\bibinfo {title} {Synthesis of uniform two-dimensional non-layered a-mnse by molecular sieves modified chemical vapor deposition},\ }\href {https://doi.org/https://doi.org/10.1016/j.rinp.2023.106321} {\bibfield  {journal} {\bibinfo  {journal} {Results in Physics}\ }\textbf {\bibinfo {volume} {47}},\ \bibinfo {pages} {106321} (\bibinfo {year} {2023})}\BibitemShut {NoStop}%
\bibitem [{\citenamefont {Milutinovi{\'{c}}}\ \emph {et~al.}(2004)\citenamefont {Milutinovi{\'{c}}}, \citenamefont {Popovi{\'{c}}}, \citenamefont {Tomi{\'{c}}},\ and\ \citenamefont {Devi{\'{c}}}}]{milutinovic2004}%
  \BibitemOpen
  \bibfield  {author} {\bibinfo {author} {\bibfnamefont {A.}~\bibnamefont {Milutinovi{\'{c}}}}, \bibinfo {author} {\bibfnamefont {Z.~V.}\ \bibnamefont {Popovi{\'{c}}}}, \bibinfo {author} {\bibfnamefont {N.}~\bibnamefont {Tomi{\'{c}}}},\ and\ \bibinfo {author} {\bibfnamefont {S.}~\bibnamefont {Devi{\'{c}}}},\ }\bibfield  {title} {\bibinfo {title} {Raman spectroscopy of polycrystalline $\alpha$ - mnse},\ }in\ \href {https://doi.org/10.4028/www.scientific.net/MSF.453-454.299} {\emph {\bibinfo {booktitle} {Progress in Advanced Materials and Processes}}},\ \bibinfo {series} {Materials Science Forum}, Vol.\ \bibinfo {volume} {453}\ (\bibinfo  {publisher} {Trans Tech Publications Ltd},\ \bibinfo {year} {2004})\ pp.\ \bibinfo {pages} {299--304}\BibitemShut {NoStop}%
\bibitem [{\citenamefont {Popovi\ifmmode~\acute{c}\else \'{c}\fi{}}\ and\ \citenamefont {Milutinovi\ifmmode~\acute{c}\else \'{c}\fi{}}(2006)}]{PhysRevB.73.155203}%
  \BibitemOpen
  \bibfield  {author} {\bibinfo {author} {\bibfnamefont {Z.~V.}\ \bibnamefont {Popovi\ifmmode~\acute{c}\else \'{c}\fi{}}}\ and\ \bibinfo {author} {\bibfnamefont {A.}~\bibnamefont {Milutinovi\ifmmode~\acute{c}\else \'{c}\fi{}}},\ }\bibfield  {title} {\bibinfo {title} {Far-infrared reflectivity and raman scattering study of $\ensuremath{\alpha}\text{\ensuremath{-}}\mathrm{Mn}\mathrm{Se}$},\ }\href {https://doi.org/10.1103/PhysRevB.73.155203} {\bibfield  {journal} {\bibinfo  {journal} {Phys. Rev. B}\ }\textbf {\bibinfo {volume} {73}},\ \bibinfo {pages} {155203} (\bibinfo {year} {2006})}\BibitemShut {NoStop}%
\bibitem [{\citenamefont {Kittel}\ \emph {et~al.}(2015)\citenamefont {Kittel}, \citenamefont {McEuen},\ and\ \citenamefont {Sons}}]{Kittel2015}%
  \BibitemOpen
  \bibfield  {author} {\bibinfo {author} {\bibfnamefont {C.}~\bibnamefont {Kittel}}, \bibinfo {author} {\bibfnamefont {P.}~\bibnamefont {McEuen}},\ and\ \bibinfo {author} {\bibfnamefont {J.~W.~.}\ \bibnamefont {Sons}},\ }\href {https://books.google.com.sg/books?id=rAMujwEACAAJ} {\emph {\bibinfo {title} {Introduction to Solid State Physics}}}\ (\bibinfo  {publisher} {John Wiley \& Sons},\ \bibinfo {year} {2015})\BibitemShut {NoStop}%
\bibitem [{\citenamefont {Pradhan}\ \emph {et~al.}(8 08)\citenamefont {Pradhan}, \citenamefont {Dalal},\ and\ \citenamefont {De}}]{Pradhan2018}%
  \BibitemOpen
  \bibfield  {author} {\bibinfo {author} {\bibfnamefont {S.~K.}\ \bibnamefont {Pradhan}}, \bibinfo {author} {\bibfnamefont {B.}~\bibnamefont {Dalal}},\ and\ \bibinfo {author} {\bibfnamefont {S.~K.}\ \bibnamefont {De}},\ }\bibfield  {title} {\bibinfo {title} {Exchange bias effect in a finite site disordered canted antiferromagnet},\ }\href {https://doi.org/10.1088/1361-648X/aad7d6} {\bibfield  {journal} {\bibinfo  {journal} {Journal of Physics: Condensed Matter}\ }\textbf {\bibinfo {volume} {30}},\ \bibinfo {pages} {365801} (\bibinfo {year} {2018-08})}\BibitemShut {NoStop}%
\bibitem [{\citenamefont {Du}\ \emph {et~al.}(2019)\citenamefont {Du}, \citenamefont {Tang}, \citenamefont {Zhao}, \citenamefont {Li}, \citenamefont {Yang}, \citenamefont {Hu}, \citenamefont {Bai}, \citenamefont {Wang}, \citenamefont {Watanabe}, \citenamefont {Taniguchi}, \citenamefont {Shi}, \citenamefont {Yu}, \citenamefont {Bai}, \citenamefont {Hasan}, \citenamefont {Zhang},\ and\ \citenamefont {Sun}}]{Du}%
  \BibitemOpen
  \bibfield  {author} {\bibinfo {author} {\bibfnamefont {L.}~\bibnamefont {Du}}, \bibinfo {author} {\bibfnamefont {J.}~\bibnamefont {Tang}}, \bibinfo {author} {\bibfnamefont {Y.}~\bibnamefont {Zhao}}, \bibinfo {author} {\bibfnamefont {X.}~\bibnamefont {Li}}, \bibinfo {author} {\bibfnamefont {R.}~\bibnamefont {Yang}}, \bibinfo {author} {\bibfnamefont {X.}~\bibnamefont {Hu}}, \bibinfo {author} {\bibfnamefont {X.}~\bibnamefont {Bai}}, \bibinfo {author} {\bibfnamefont {X.}~\bibnamefont {Wang}}, \bibinfo {author} {\bibfnamefont {K.}~\bibnamefont {Watanabe}}, \bibinfo {author} {\bibfnamefont {T.}~\bibnamefont {Taniguchi}}, \bibinfo {author} {\bibfnamefont {D.}~\bibnamefont {Shi}}, \bibinfo {author} {\bibfnamefont {G.}~\bibnamefont {Yu}}, \bibinfo {author} {\bibfnamefont {X.}~\bibnamefont {Bai}}, \bibinfo {author} {\bibfnamefont {T.}~\bibnamefont {Hasan}}, \bibinfo {author} {\bibfnamefont {G.}~\bibnamefont {Zhang}},\ and\ \bibinfo {author} {\bibfnamefont {Z.}~\bibnamefont {Sun}},\ }\bibfield  {title} {\bibinfo
  {title} {Lattice dynamics, phonon chirality, and spin–phonon coupling in 2d itinerant ferromagnet fe3gete2},\ }\href {https://doi.org/https://doi.org/10.1002/adfm.201904734} {\bibfield  {journal} {\bibinfo  {journal} {Advanced Functional Materials}\ }\textbf {\bibinfo {volume} {29}},\ \bibinfo {pages} {1904734} (\bibinfo {year} {2019})},\ \Eprint {https://arxiv.org/abs/https://onlinelibrary.wiley.com/doi/pdf/10.1002/adfm.201904734} {https://onlinelibrary.wiley.com/doi/pdf/10.1002/adfm.201904734} \BibitemShut {NoStop}%
\bibitem [{\citenamefont {Bera}\ \emph {et~al.}(5 01)\citenamefont {Bera}, \citenamefont {Sarathi~Rana}, \citenamefont {Kalyan~Pradhan}, \citenamefont {Palit}, \citenamefont {Saha}, \citenamefont {Kalimuddin}, \citenamefont {Bera}, \citenamefont {Debnath}, \citenamefont {Das}, \citenamefont {Singha~Roy}, \citenamefont {Datta},\ and\ \citenamefont {Mondal}}]{Bera2025}%
  \BibitemOpen
  \bibfield  {author} {\bibinfo {author} {\bibfnamefont {A.}~\bibnamefont {Bera}}, \bibinfo {author} {\bibfnamefont {P.}~\bibnamefont {Sarathi~Rana}}, \bibinfo {author} {\bibfnamefont {S.}~\bibnamefont {Kalyan~Pradhan}}, \bibinfo {author} {\bibfnamefont {M.}~\bibnamefont {Palit}}, \bibinfo {author} {\bibfnamefont {S.}~\bibnamefont {Saha}}, \bibinfo {author} {\bibfnamefont {S.}~\bibnamefont {Kalimuddin}}, \bibinfo {author} {\bibfnamefont {S.}~\bibnamefont {Bera}}, \bibinfo {author} {\bibfnamefont {T.}~\bibnamefont {Debnath}}, \bibinfo {author} {\bibfnamefont {S.}~\bibnamefont {Das}}, \bibinfo {author} {\bibfnamefont {D.}~\bibnamefont {Singha~Roy}}, \bibinfo {author} {\bibfnamefont {S.}~\bibnamefont {Datta}},\ and\ \bibinfo {author} {\bibfnamefont {M.}~\bibnamefont {Mondal}},\ }\bibfield  {title} {\bibinfo {title} {Raman signatures of inversion symmetry breaking structural transition in quasi-1d compound, (tase4)3i},\ }\href {https://doi.org/10.1088/1361-648X/ada843} {\bibfield  {journal} {\bibinfo  {journal}
  {Journal of Physics: Condensed Matter}\ }\textbf {\bibinfo {volume} {37}},\ \bibinfo {pages} {125403} (\bibinfo {year} {2025-01})}\BibitemShut {NoStop}%
\bibitem [{\citenamefont {Balkanski}\ \emph {et~al.}(1975)\citenamefont {Balkanski}, \citenamefont {Jain}, \citenamefont {Beserman},\ and\ \citenamefont {Jouanne}}]{PhysRevB.12.4328}%
  \BibitemOpen
  \bibfield  {author} {\bibinfo {author} {\bibfnamefont {M.}~\bibnamefont {Balkanski}}, \bibinfo {author} {\bibfnamefont {K.~P.}\ \bibnamefont {Jain}}, \bibinfo {author} {\bibfnamefont {R.}~\bibnamefont {Beserman}},\ and\ \bibinfo {author} {\bibfnamefont {M.}~\bibnamefont {Jouanne}},\ }\bibfield  {title} {\bibinfo {title} {Theory of interference distortion of raman scattering line shapes in semiconductors},\ }\href {https://doi.org/10.1103/PhysRevB.12.4328} {\bibfield  {journal} {\bibinfo  {journal} {Phys. Rev. B}\ }\textbf {\bibinfo {volume} {12}},\ \bibinfo {pages} {4328} (\bibinfo {year} {1975})}\BibitemShut {NoStop}%
\bibitem [{\citenamefont {Casto}\ \emph {et~al.}(2015)\citenamefont {Casto}, \citenamefont {Clune}, \citenamefont {Yokosuk}, \citenamefont {Musfeldt}, \citenamefont {Williams}, \citenamefont {Zhuang}, \citenamefont {Lin}, \citenamefont {Xiao}, \citenamefont {Hennig}, \citenamefont {Sales}, \citenamefont {Yan},\ and\ \citenamefont {Mandrus}}]{Casto2015}%
  \BibitemOpen
  \bibfield  {author} {\bibinfo {author} {\bibfnamefont {L.~D.}\ \bibnamefont {Casto}}, \bibinfo {author} {\bibfnamefont {A.~J.}\ \bibnamefont {Clune}}, \bibinfo {author} {\bibfnamefont {M.~O.}\ \bibnamefont {Yokosuk}}, \bibinfo {author} {\bibfnamefont {J.~L.}\ \bibnamefont {Musfeldt}}, \bibinfo {author} {\bibfnamefont {T.~J.}\ \bibnamefont {Williams}}, \bibinfo {author} {\bibfnamefont {H.~L.}\ \bibnamefont {Zhuang}}, \bibinfo {author} {\bibfnamefont {M.-W.}\ \bibnamefont {Lin}}, \bibinfo {author} {\bibfnamefont {K.}~\bibnamefont {Xiao}}, \bibinfo {author} {\bibfnamefont {R.~G.}\ \bibnamefont {Hennig}}, \bibinfo {author} {\bibfnamefont {B.~C.}\ \bibnamefont {Sales}}, \bibinfo {author} {\bibfnamefont {J.-Q.}\ \bibnamefont {Yan}},\ and\ \bibinfo {author} {\bibfnamefont {D.}~\bibnamefont {Mandrus}},\ }\bibfield  {title} {\bibinfo {title} {Strong spin-lattice coupling in crsite3},\ }\href {https://doi.org/10.1063/1.4914134} {\bibfield  {journal} {\bibinfo  {journal} {APL Materials}\ }\textbf {\bibinfo {volume}
  {3}},\ \bibinfo {pages} {041515} (\bibinfo {year} {2015})},\ \Eprint {https://arxiv.org/abs/https://pubs.aip.org/aip/apm/article-pdf/doi/10.1063/1.4914134/14558234/041515\_1\_online.pdf} {https://pubs.aip.org/aip/apm/article-pdf/doi/10.1063/1.4914134/14558234/041515\_1\_online.pdf} \BibitemShut {NoStop}%
\bibitem [{\citenamefont {Ghosh}\ \emph {et~al.}(2021)\citenamefont {Ghosh}, \citenamefont {Palit}, \citenamefont {Maity}, \citenamefont {Dwij}, \citenamefont {Rana},\ and\ \citenamefont {Datta}}]{Ghosh2021}%
  \BibitemOpen
  \bibfield  {author} {\bibinfo {author} {\bibfnamefont {A.}~\bibnamefont {Ghosh}}, \bibinfo {author} {\bibfnamefont {M.}~\bibnamefont {Palit}}, \bibinfo {author} {\bibfnamefont {S.}~\bibnamefont {Maity}}, \bibinfo {author} {\bibfnamefont {V.}~\bibnamefont {Dwij}}, \bibinfo {author} {\bibfnamefont {S.}~\bibnamefont {Rana}},\ and\ \bibinfo {author} {\bibfnamefont {S.}~\bibnamefont {Datta}},\ }\bibfield  {title} {\bibinfo {title} {Spin-phonon coupling and magnon scattering in few-layer antiferromagnetic ${\mathrm{feps}}_{3}$},\ }\href {https://doi.org/10.1103/PhysRevB.103.064431} {\bibfield  {journal} {\bibinfo  {journal} {Phys. Rev. B}\ }\textbf {\bibinfo {volume} {103}},\ \bibinfo {pages} {064431} (\bibinfo {year} {2021})}\BibitemShut {NoStop}%
\bibitem [{\citenamefont {Lee}\ \emph {et~al.}(2010)\citenamefont {Lee}, \citenamefont {Fang}, \citenamefont {Vlahos}, \citenamefont {Ke}, \citenamefont {Jung}, \citenamefont {Kourkoutis}, \citenamefont {Kim}, \citenamefont {Ryan}, \citenamefont {Heeg}, \citenamefont {Roeckerath}, \citenamefont {Goian}, \citenamefont {Bernhagen}, \citenamefont {Uecker}, \citenamefont {Hammel}, \citenamefont {Rabe}, \citenamefont {Kamba}, \citenamefont {Schubert}, \citenamefont {Freeland}, \citenamefont {Muller}, \citenamefont {Fennie}, \citenamefont {Schiffer}, \citenamefont {Gopalan}, \citenamefont {Johnston-Halperin},\ and\ \citenamefont {Schlom}}]{Lee}%
  \BibitemOpen
  \bibfield  {author} {\bibinfo {author} {\bibfnamefont {J.~H.}\ \bibnamefont {Lee}}, \bibinfo {author} {\bibfnamefont {L.}~\bibnamefont {Fang}}, \bibinfo {author} {\bibfnamefont {E.}~\bibnamefont {Vlahos}}, \bibinfo {author} {\bibfnamefont {X.}~\bibnamefont {Ke}}, \bibinfo {author} {\bibfnamefont {Y.~W.}\ \bibnamefont {Jung}}, \bibinfo {author} {\bibfnamefont {L.~F.}\ \bibnamefont {Kourkoutis}}, \bibinfo {author} {\bibfnamefont {J.-W.}\ \bibnamefont {Kim}}, \bibinfo {author} {\bibfnamefont {P.~J.}\ \bibnamefont {Ryan}}, \bibinfo {author} {\bibfnamefont {T.}~\bibnamefont {Heeg}}, \bibinfo {author} {\bibfnamefont {M.}~\bibnamefont {Roeckerath}}, \bibinfo {author} {\bibfnamefont {V.}~\bibnamefont {Goian}}, \bibinfo {author} {\bibfnamefont {M.}~\bibnamefont {Bernhagen}}, \bibinfo {author} {\bibfnamefont {R.}~\bibnamefont {Uecker}}, \bibinfo {author} {\bibfnamefont {P.~C.}\ \bibnamefont {Hammel}}, \bibinfo {author} {\bibfnamefont {K.~M.}\ \bibnamefont {Rabe}}, \bibinfo {author} {\bibfnamefont {S.}~\bibnamefont
  {Kamba}}, \bibinfo {author} {\bibfnamefont {J.}~\bibnamefont {Schubert}}, \bibinfo {author} {\bibfnamefont {J.~W.}\ \bibnamefont {Freeland}}, \bibinfo {author} {\bibfnamefont {D.~A.}\ \bibnamefont {Muller}}, \bibinfo {author} {\bibfnamefont {C.~J.}\ \bibnamefont {Fennie}}, \bibinfo {author} {\bibfnamefont {P.}~\bibnamefont {Schiffer}}, \bibinfo {author} {\bibfnamefont {V.}~\bibnamefont {Gopalan}}, \bibinfo {author} {\bibfnamefont {E.}~\bibnamefont {Johnston-Halperin}},\ and\ \bibinfo {author} {\bibfnamefont {D.~G.}\ \bibnamefont {Schlom}},\ }\bibfield  {title} {\bibinfo {title} {A strong ferroelectric ferromagnet created by means of spin-lattice coupling},\ }\href {https://doi.org/10.1038/nature09331} {\bibfield  {journal} {\bibinfo  {journal} {Nature}\ }\textbf {\bibinfo {volume} {466}},\ \bibinfo {pages} {954} (\bibinfo {year} {2010})}\BibitemShut {NoStop}%
\bibitem [{\citenamefont {Iliev}\ \emph {et~al.}(2007{\natexlab{a}})\citenamefont {Iliev}, \citenamefont {Abrashev}, \citenamefont {Litvinchuk}, \citenamefont {Hadjiev}, \citenamefont {Guo},\ and\ \citenamefont {Gupta}}]{PhysRevB.75.104118}%
  \BibitemOpen
  \bibfield  {author} {\bibinfo {author} {\bibfnamefont {M.~N.}\ \bibnamefont {Iliev}}, \bibinfo {author} {\bibfnamefont {M.~V.}\ \bibnamefont {Abrashev}}, \bibinfo {author} {\bibfnamefont {A.~P.}\ \bibnamefont {Litvinchuk}}, \bibinfo {author} {\bibfnamefont {V.~G.}\ \bibnamefont {Hadjiev}}, \bibinfo {author} {\bibfnamefont {H.}~\bibnamefont {Guo}},\ and\ \bibinfo {author} {\bibfnamefont {A.}~\bibnamefont {Gupta}},\ }\bibfield  {title} {\bibinfo {title} {Raman spectroscopy of ordered double perovskite ${\mathrm{la}}_{2}\mathrm{Co}\mathrm{Mn}{\mathrm{o}}_{6}$ thin films},\ }\href {https://doi.org/10.1103/PhysRevB.75.104118} {\bibfield  {journal} {\bibinfo  {journal} {Phys. Rev. B}\ }\textbf {\bibinfo {volume} {75}},\ \bibinfo {pages} {104118} (\bibinfo {year} {2007}{\natexlab{a}})}\BibitemShut {NoStop}%
\bibitem [{\citenamefont {Iliev}\ \emph {et~al.}(2007{\natexlab{b}})\citenamefont {Iliev}, \citenamefont {Guo},\ and\ \citenamefont {Gupta}}]{Iliev}%
  \BibitemOpen
  \bibfield  {author} {\bibinfo {author} {\bibfnamefont {M.~N.}\ \bibnamefont {Iliev}}, \bibinfo {author} {\bibfnamefont {H.}~\bibnamefont {Guo}},\ and\ \bibinfo {author} {\bibfnamefont {A.}~\bibnamefont {Gupta}},\ }\bibfield  {title} {\bibinfo {title} {Raman spectroscopy evidence of strong spin-phonon coupling in epitaxial thin films of the double perovskite la2nimno6},\ }\href {https://doi.org/10.1063/1.2721142} {\bibfield  {journal} {\bibinfo  {journal} {Applied Physics Letters}\ }\textbf {\bibinfo {volume} {90}},\ \bibinfo {pages} {151914} (\bibinfo {year} {2007}{\natexlab{b}})},\ \Eprint {https://arxiv.org/abs/https://pubs.aip.org/aip/apl/article-pdf/doi/10.1063/1.2721142/13172034/151914\_1\_online.pdf} {https://pubs.aip.org/aip/apl/article-pdf/doi/10.1063/1.2721142/13172034/151914\_1\_online.pdf} \BibitemShut {NoStop}%
\bibitem [{\citenamefont {Ghorai}\ \emph {et~al.}(2024)\citenamefont {Ghorai}, \citenamefont {Ghosh}, \citenamefont {Patra}, \citenamefont {Samal}, \citenamefont {Senapati},\ and\ \citenamefont {Sahoo}}]{Ghorai2024}%
  \BibitemOpen
  \bibfield  {author} {\bibinfo {author} {\bibfnamefont {G.}~\bibnamefont {Ghorai}}, \bibinfo {author} {\bibfnamefont {K.}~\bibnamefont {Ghosh}}, \bibinfo {author} {\bibfnamefont {A.}~\bibnamefont {Patra}}, \bibinfo {author} {\bibfnamefont {P.}~\bibnamefont {Samal}}, \bibinfo {author} {\bibfnamefont {K.}~\bibnamefont {Senapati}},\ and\ \bibinfo {author} {\bibfnamefont {P.~K.}\ \bibnamefont {Sahoo}},\ }\href {https://arxiv.org/abs/2403.04426} {\bibinfo {title} {Spin-phonon interaction in quasi 2d- cr$_2te_3$}} (\bibinfo {year} {2024}),\ \Eprint {https://arxiv.org/abs/2403.04426} {arXiv:2403.04426 [cond-mat.mtrl-sci]} \BibitemShut {NoStop}%
\bibitem [{\citenamefont {Wu}(2012)}]{Wu2012}%
  \BibitemOpen
  \bibfield  {author} {\bibinfo {author} {\bibfnamefont {S.~Y.}\ \bibnamefont {Wu}},\ }\bibfield  {title} {\bibinfo {title} {Application of raman microscopy for spin-phonon coupling and magnon excitation study in nanocrystals}\ }(\bibinfo {year} {2012})\BibitemShut {NoStop}%
\bibitem [{\citenamefont {Lockwood}\ \emph {et~al.}(1983)\citenamefont {Lockwood}, \citenamefont {Katiyar},\ and\ \citenamefont {So}}]{Lockwood1983}%
  \BibitemOpen
  \bibfield  {author} {\bibinfo {author} {\bibfnamefont {D.~J.}\ \bibnamefont {Lockwood}}, \bibinfo {author} {\bibfnamefont {R.~S.}\ \bibnamefont {Katiyar}},\ and\ \bibinfo {author} {\bibfnamefont {V.~C.~Y.}\ \bibnamefont {So}},\ }\bibfield  {title} {\bibinfo {title} {${B}_{1g}$ mode softening in fe${\mathrm{f}}_{2}$},\ }\href {https://doi.org/10.1103/PhysRevB.28.1983} {\bibfield  {journal} {\bibinfo  {journal} {Phys. Rev. B}\ }\textbf {\bibinfo {volume} {28}},\ \bibinfo {pages} {1983} (\bibinfo {year} {1983})}\BibitemShut {NoStop}%
\bibitem [{\citenamefont {Granado}\ \emph {et~al.}(1999)\citenamefont {Granado}, \citenamefont {Garc\'{\i}a}, \citenamefont {Sanjurjo}, \citenamefont {Rettori}, \citenamefont {Torriani}, \citenamefont {Prado}, \citenamefont {S\'anchez}, \citenamefont {Caneiro},\ and\ \citenamefont {Oseroff}}]{GranadoE}%
  \BibitemOpen
  \bibfield  {author} {\bibinfo {author} {\bibfnamefont {E.}~\bibnamefont {Granado}}, \bibinfo {author} {\bibfnamefont {A.}~\bibnamefont {Garc\'{\i}a}}, \bibinfo {author} {\bibfnamefont {J.~A.}\ \bibnamefont {Sanjurjo}}, \bibinfo {author} {\bibfnamefont {C.}~\bibnamefont {Rettori}}, \bibinfo {author} {\bibfnamefont {I.}~\bibnamefont {Torriani}}, \bibinfo {author} {\bibfnamefont {F.}~\bibnamefont {Prado}}, \bibinfo {author} {\bibfnamefont {R.~D.}\ \bibnamefont {S\'anchez}}, \bibinfo {author} {\bibfnamefont {A.}~\bibnamefont {Caneiro}},\ and\ \bibinfo {author} {\bibfnamefont {S.~B.}\ \bibnamefont {Oseroff}},\ }\bibfield  {title} {\bibinfo {title} {Magnetic ordering effects in the raman spectra of ${\mathrm{la}}_{1\ensuremath{-}x}{\mathrm{mn}}_{1\ensuremath{-}x}{\mathrm{o}}_{3}$},\ }\href {https://doi.org/10.1103/PhysRevB.60.11879} {\bibfield  {journal} {\bibinfo  {journal} {Phys. Rev. B}\ }\textbf {\bibinfo {volume} {60}},\ \bibinfo {pages} {11879} (\bibinfo {year} {1999})}\BibitemShut {NoStop}%
\end{thebibliography}
%

\end{document}